\documentclass[]{emulateapj}
\begin{document}

\newcommand{\neff}{$N_{\rm eff}$}
\newcommand{\mneff}{N_{\rm eff}}
\newcommand{\siew}{$W_{1526}$}
\newcommand{\lya}{Ly$\alpha$}
\newcommand{\lyb}{Ly$\beta$}
\newcommand{\zabs}{z$_{abs}$}
\newcommand{\mnhi}{N_{\rm HI}}
\newcommand{\zem}{z$_{em}$}
\newcommand{\ncol}{$\log$ $N_{HI}$}
\newcommand{\atomspercm}{atoms cm$^{-2}$}
\newcommand{\nhi}{$N_{\rm HI}$}
\newcommand{\mkms}{{\rm \; km\;s^{-1}}}
\newcommand{\cm}[1]{\, {\rm cm^{#1}}}
\newcommand{\N}[1]{{N({\rm #1})}}
\newcommand{\w}{$W_{1526}$} %EW for SI II 1526
\newcommand{\deltaalpha}{$\Delta\alpha$}
\newcommand{\mdeltaalpha}{\Delta\alpha}
\newcommand{\rAA}{{\rm \AA}}
\def\ltp{\left ( \,}
\def\rtp{\, \right  ) }

\title{H I Column Densities, Metallicities, and Dust Extinction of 
Metal-Strong Damped \lya\ Systems\altaffilmark{1}}

\author{
 Kyle F. Kaplan\altaffilmark{2},
 J. Xavier Prochaska\altaffilmark{2,3},
 St\'ephane Herbert-Fort\altaffilmark{4},
 Sara L. Ellison\altaffilmark{5},
 Miroslava Dessauges-Zavadsky\altaffilmark{6}
}

\altaffiltext{1}{Observations reported here were obtained at the MMT Observatory, 
a joint facility of the University of Arizona and the Smithsonian Institution.}
\altaffiltext{2}{Department of Astronomy and Astrophysics,
University of California, Santa Cruz, Santa Cruz, CA 95064} 
\altaffiltext{3}{University of California Observatories - 
Lick Observatory, University of California, Santa Cruz, CA 95064} 
\altaffiltext{4}{University of Arizona/Steward Observatory, 
933 N Cherry Avenue, Tucson, AZ 85721}
\altaffiltext{5}{Department of Physics and Astronomy, University of Victoria,
Victoria, British Columbia, V8P 1A1, Canada.}
\altaffiltext{6}{Observatoire de Gen\`eve, 51 Ch. des Maillettes, 
1290 Sauverny, Switzerland}

\begin{abstract}
With the Blue Channel Spectrograph (BCS) on the MMT telescope,
we have obtained spectra to the atmospheric
cutoff of quasars previously known to show at least one absorption
system at $z>1.6$ with very strong metal lines.   We refer
to these absorbers as candidate metal-strong damped \lya\ systems
(cMSDLAs), the majority of which were culled from the Sloan Digital
Sky Survey.  The BCS/MMT spectra yield precise estimates of the
\ion{H}{1} column densities (\nhi) of the systems through Voigt
profile analysis of their \lya\ transitions.  Nearly all of the
cMSDLAs (41/43) satisfy the \nhi\ criterion of DLAs, $10^{20.3}$ \atomspercm.
As a population, these systems have systematically higher \nhi\ values
than DLAs chosen randomly from quasar sightlines.  Combining 
our \nhi\ measurements with previously measured metal column densities,
we estimate metallicities for the MSDLAs.  These systems have significantly
higher values than randomly selected DLAs; at $z \approx 2$, the
MSDLAs show a median metallicity [M/H]~$\approx -0.67$ that is $0.6$\,dex
higher than a corresponding control sample. 
This establishes MSDLAs as having
amongst the most metal-rich gas in the high $z$ universe.
Our measurements extend the observed correlation between \ion{Si}{2}~1526
equivalent width and the gas metallicity to higher values.  If
interpreted as a mass-metallicity relation, this implies the MSDLAs
are the high mass subset of the DLA population.  We demonstrate
that dust in the MSDLAs reddens their background quasars, with a
median shift in the spectral slope of $\delta\alpha = 0.29$.
Assuming an SMC extinction law, this implies a median
reddening $E_{B-V} \approx 0.025$\,mag and visual extinction 
$A_V \approx 0.076$\,mag.  The latter quantity yields a 
dust-to-gas ratio of $log(A_V / N_{\rm HI}) \approx -22.0$, 
very similar to estimation for the SMC.
Future studies of MSDLAs offer the opportunity to study the extinction,
nucleosynthesis, and kinematics of the most chemically evolved, gas-rich
galaxies at high $z$.
\end{abstract}

\section{Introduction}
%\tau and \Tau

A Damped \lya\ Absorption System (DLA) is defined as a neutral 
hydrogen absorber along a Quasi-Stellar Object (QSO) sight line with an \ion{H}{1} 
column density of $\mnhi \ge 10^{20.3}$ \atomspercm\ 
\citep[see][for a review]{wgp05}.  
DLAs provide an important way to study the gas and dust content 
of the early universe \citep[e.g.][]{pshk94,lu96,pw01}.  
%The heavy elements that make up the gas and dust in DLAs were 
%not created in Big Bang nucleosynthesis.  
Their heavy elements were 
synthesized either in nuclear reactions occurring in stellar cores 
or when massive stars go supernova.  Studies of 
DLAs, therefore, probe the processes of stellar evolution in young galaxies.
%The absorption lines of neutral hydrogen and metals in the DLA system
%are imprinted in the QSO's spectrum.  
The large surface density of neutral hydrogen that defines DLAs
means these systems contain most of the star forming gas 
in the early universe \citep{phw05,opb+07}.  
Since the presence of heavy metals in DLAs implies the existence of stars, 
DLAs likely trace high redshift galaxies along the sight line between the QSO and 
earth.  These high redshift galaxies are thought 
to be precursors of galaxies like the Milky Way
%Find which paper nagamine is supposed to be
\citep[e.g.][]{nsh04,pgp+08}.  
%This assertion is supported by the imaging of galaxies with 
%redshifts matching DLAs \citep{warren01, mff04}, although
%it is difficult to conclude much about the nature and morphology of 
%the larger population of high redshift DLAs from this small sample of images. 
%although they appear to be galaxies, or at least galactic precursors.

A subset of DLAs exhibit
especially strong metal absorption lines.
The metal-strong damped \lya\ systems (MSDLAs)
enable analysis of tens of elements in individual galaxies \citep{phw03}. 
The metal-strong criteria were defined by 
\cite[][; hereafter, H06]{shf06} to be absorbers showing 
ionic column densities that satisfy
$\log$ N(Zn$^+$)$\ge$13.15 and/or $\log$ N(Si$^+$)$\ge$15.95.  
The MSDLAs, therefore, are absorbers that satisfy the metal-strong
criteria and also the DLA criterion on \nhi.
H06 identified a number of MSDLA candidates from the 
Sloan Digital Sky Survey (SDSS) and presented higher resolution follow-up 
spectra of the metal-line transitions.   The strong metal absorption lines of these candidates suggests a high metallicity.
For many of these MSDLA candidates, however, the \lya\ absorption line 
was bluer than the wavelength coverage in the SDSS spectra, precluding measurements of their H I column densities.  
Therefore, these absorption systems could not be classified as damped
\lya\ systems nor could the authors provide metallicity estimates for
the gas.

%INSERT SENTENCE OR TWO (PROBABLY WITH REFERENCE[S] ABOUT THE NATURE OF DLAS, ARE THEY DWARF GALAXIES?
%JXP The following paragraph was good for you but not necessary in a paper like this.
%The \lya\ spectral line is the first line in the hydrogen Lyman series. It normally lies in the ultraviolet part of the electromagnetic spectrum (at $\lambda$=1215.67 \AA) which is opaque to the atmosphere and not visible to observers on the ground.  DLAs at high enough redshift (\zabs$\gtrsim$1.5) are displaced from their rest wavelengths into the visible spectrum allowing them to be studied here on Earth from the ground. To study DLAs at lower redshifts would require space based telescopes.  Astronomers are presented with the unique opportunity of being able to study the gas in the interstellar medium of metal enriched galaxies at these high redshifts.
The wavelengths covered by SDSS spectra range from 3800 to 9200 \AA\ \citep{sdssdr6}.  This range allows for the identification of DLAs with redshifts corresponding to 
$z_{\rm abs} \gtrsim 2.2$.  
For absorbers with 1.5 $\lesssim$ \zabs\ $\lesssim$ 2.2,  
the \lya\ absorption line lies blueward of the SDSS spectral coverage,
but these systems can be observed from ground-based observatories 
equipped with blue-sensitive spectrometers.  
With this observational goal in mind,
we obtained new spectra from the MMT telescope of various QSO sight lines containing MSDLA candidates using the Blue Channel Spectrograph (BCS).  

%The first step in our analysis is determining whether these MSDLA candidates are actually DLAs.  We detail this in INSERT SECTION NAME HERE.  With wavelength coverage of the absorbers' \lya\ line, a Voigt profile is fit to the line to determine the column density of neutral hydrogen.  If the column density of neutral hydrogen is (\ncol$\ge$20.3 \atomspercm), then the absorber is classified as a DLA.  With the neutral hydrogen column density measured and metal column densities previously determined from high resolution spectra, the metallicity [M/H] of each absorber can be calculated.   The measurements of metal column densities and the calculation of metallicities are covered in section INSERT SECTION NAME HERE. 

%{\bf [The remaining paragraphs of the Intro
%are good, but reduce them by about 50\%.  Don't insert results from
%this paper nor reference the section numbers except for by finishing 
%Intro with a short paragraph that outlines the paper (see ones
%I have written as an example)]}

%Summarization of classification scheme.  I have placed it here in the intro, but it might be moved elsewhere.  Hopefully this will clear up the confusion on how things are classified in this paper.
Here we summarize our classification scheme.  Systems with $\mnhi \ge 10^{20.3}$ \atomspercm\ are classified as DLAs.  Systems with  H I column densities meeting the DLA criteria  stated above and with measured metal column densities of $\log$ N(Zn$^+$)$\ge$13.15 and/or $\log$ N(Si$^+$)$\ge$15.95 are classified as MSDLAs.  Candidate MSDLAs are systems identified with strong metal lines in the SDSS spectra (or elsewhere for a few systems) for which there was previously no \lya\ coverage
and/or no precise measurement of the metal column density.  
We refer to these candidate MSDLAs as cMSDLAs.
Each cMSDLA has had its spectrum taken with the MMT/BCS in order to obtain \lya\ coverage.  
%Some QSO sightlines contain more than one absorption system, some of which meet the DLA criteria.  We are not interested in these systems and they are not classified as candidate MSDLAs.

%Spectral absorption lines show that Metal Strong DLAs contain heavy elements in the gas phase.  
The gas identified with MSDLAs is analogous to the
%interstellar gas clouds 
HI regions
found in the Milky Way \citep[e.g.][]{ss96}.  
Several heavy elements detected in the gas of MSDLAs are also known to make up interstellar dust in the Milky Way and other nearby galaxies.
 %Do I have to cite anything here about milky way and nearby galaxy dust?
Dust in the Milky Way is comprised of heavy 
elements including 
C, O, Mg, Ni, S, K, Mn, Si, Fe, Al, Ti, and Ca 
\citep[e.g.][]{savage+mathis79,jenkins09}.  
\cite{shf06} presents the abundance ratios of previously observed MSDLAs showing their ratios approach solar.
Dust particles preferentially absorb short wavelengths of light, reddening the color of background objects.  
\cite{ehl05} have performed the only extinction estimate
for a complete survey of DLAs and set an upper limit to their
average reddening of $E_{B-V} < 0.04$\, mag.
Reddening has been studied in the low redshift universe where nearby galaxies can be well resolved. 
In contrast, the properties of dust in high redshift galaxies is very
poorly constrained.
QSO absorption systems provide a probe for studying reddening in high redshift galaxies due to absorption of the QSO's light by the dust in the absorber.  
A series of studies have been performed on the extinction properties of DLAs
beginning several decades ago \citep[e.g.][]{oh84,pfb91}.
More recently, \cite{ml04} leveraged the large dataset afforded by SDSS
to examine reddening from $\approx 100$ DLAs in the 
SDSS-DR2 using the spectrophotometric observations.  
Their results show no conclusive evidence for reddening ($E_{B-V} < 0.02$ mag). 
Reddening in DLAs has also been investigated using the SDSS photometry
by \cite{vcl+06,vpw08}, who report evidence for dust.  
%\cite{vcl+06} claims to have detected reddening in 5 out of 13 DLAs while 
\cite{vpw08} found evidence for reddening when comparing the colors of
quasars background to a large sample of DLAs 
to a control sample of QSOs without DLAs.  
These authors estitimate an average reddening 
$E_{B-V} \approx 0.006$\,mag.
Given their large metal column densities, MSDLAs may be expected to show
significant reddening.
%Candidate MSDLAs contain heavy elements in the gas phase, so we hypothesize that evidence for dust can be found by detecting reddening spectrophotometrically.
%[Summarize their result more]
%[Transition to the final paragraph better] 

%Summary of paper
In this paper we first determine if our candidate MSDLAs meet the DLA and/or the 
metal-strong criteria and then we search for reddening to see if they contain dust.  We cover the observations and data reduction of our candidate MSDLA sample in Section \ref{sect:obs}.   The measurements for redshift and H I column density are given in  \textsection\ \ref{sect:nhi}.  We determine metallicities for the cMSDLAs and explore the relationship between [M/H] and the EW of the Si II 1256 line in \textsection\ \ref{sect:metal}.  Section \ref{sect:dust} covers the search for dust reddening and extinction in our data sample and presents the results.  We conclude in $\S$~\ref{sec:summary} with a brief summary.

\section{Observations and Reduction}\label{sect:obs}
We used the following criteria to select our MSDLA candidates for MMT/BCS 
spectroscopy, hereafter referred to as cMSDLAs.
We selected all SDSS-DR5 quasars with magnitudes $17 < r < 20$  
showing `strong' or
`very strong' Zn\,II 2026 or Si\,II 1808 absorption profiles  
at absorption redshifts $1.6 < z_{abs} < 2.2$. 
`Strong' is where the minimum depth of the absorption profile is $\approx 90\%$ the continuum flux,
while `very strong'  is when the minimum depth is $\le 85\%$ the continuum flux.
These designations come from visual inspection of the SDSS
spectra by JXP and SHF and correspond to the systems with 
flags 4 for `strong' or 5 for `very strong' in the tables of \cite{shf06}.
The minimum $z_{abs}$ of the selection is set  
to $\sim1.6$ by the
blue atmospheric cutoff, so that \lya\ remains accessible at  
optical wavelengths, while
the maximum is set to 2.2 because \lya\ becomes accessible in SDSS
data directly at higher redshifts.
We then gave priority to systems satisfying the metal-strong criterion
absorption in higher-resolution
follow-up Keck ESI and HIRES spectra (still lacking \lya\  
coverage), as well as to systems
without follow-up data but showing particularly strong metal  
absorption in the SDSS
discovery spectra (at redshifts $1.6 < z_{abs} < 2.2$).

Data on cMSDLAs were acquired over the course of four nights
(Sept. 17 2006, Dec. 15 2006, Dec. 16 2006, and Mar. 26 2007)  
using the MMT telescope at the MMT Observatory 
on Mt.\ Hopkins, Arizona.  
The data were taken with the MMT/BCS using the 
800 grooves/mm grating.
The slit was 1.25$''$ wide giving a 
spectral resolution of FWHM~$\approx 220 \mkms$.  
The wavelengths for these observations range from $\sim 3100$ to 5100 \AA, 
corresponding to coverage of the \lya\ transition for redshifts 
1.55$\lesssim$\zabs$\lesssim$3.1.   
The observations, totalling 41 QSOs, 
are summarized in Table~\ref{tab:observations}.  

\begin{deluxetable*}{rrrrrrr}
\tablewidth{0pc}
\tabletypesize{\footnotesize}
\tablecaption{List of MMT observations \label{tab:observations}}
\tablehead{
\colhead{QSO}&
\colhead{${RA \atop J2000}$}&
\colhead{${DEC \atop J2000}$}&
\colhead{${Date \atop (UT)}$}&
\colhead{${Exposure \atop (s)}$}&
\colhead{z$_{em}$}&
\colhead{${r \atop (Mag.)}$}
}
\startdata
J0008-0958 & 00:08:15.33 & -09:58:54.0 & 2006 Sep 17 & 1000 & 2.595 & 18.38 \\
J0016-0012 & 00:16:02.40 & -00:12:24.9 & 2006 Dec 14 &  900 & 2.087 & 18.03 \\
J0020+1534 & 00:20:28.97 & +15:34:35.9 & 2006 Sep 17 & 1000 & 1.763 & 18.79 \\
J0044+0018 & 00:44:39.32 & +00:18:22.7 & 2006 Dec 14 & 1200 & 1.868 & 18.20 \\
J0058+0115 & 00:58:14.31 & +01:15:30.3 & 2006 Dec 14 &  600 & 2.495 & 17.69 \\
J0120+1324 & 01:20:20.36 & +13:24:33.6 & 2006 Sep 17 & 1000 & 2.567 & 19.23 \\
Q0201+36   & 02:04:55.60 & +36:49:18.0 & 2006 Sep 17 & 1000 & 2.912 & 17.4  \\
J0316+0040 & 03:16:09.84 & +00:40:43.2 & 2006 Dec 14 & 1300 & 2.921 & 18.66 \\
J0755+2342 & 07:55:37.22 & +23:42:04.7 & 2006 Dec 16 &  900 & 1.825 & 17.20 \\
J0756+1648 & 07:56:36.73 & +16:48:50.7 & 2006 Dec 16 & 1200 & 2.866 & 18.76 \\
J0812+3208 & 08:12:40.68 & +32:08:08.6 & 2006 Dec 16 &  600 & 2.704 & 17.46 \\
J0820+0819 & 08:20:58.37 & +08:19:48.0 & 2006 Dec 16 &  900 & 2.519 & 18.36 \\
J0831+4025 & 08:31:08.01 & +40:25:31.0 & 2006 Dec 16 &  900 & 2.330 & 18.87 \\
J0840+4942 & 08:40:32.95 & +49:42:52.8 & 2006 Dec 16 & 1200 & 2.076 & 19.02 \\
J0856+3350 & 08:56:18.23 & +33:50:42.8 & 2006 Dec 16 &  450 & 1.726 & 17.18 \\
J0912-0047 & 09:12:47.59 & -00:47:17.4 & 2006 Dec 16 &  900 & 2.859 & 18.68 \\
J0927+5823 & 09:27:08.88 & +58:23:19.4 & 2006 Dec 16 &  900 & 1.910 & 18.27 \\
J0938+3805 & 09:38:46.77 & +38:05:49.8 & 2006 Dec 16 &  300 & 1.827 & 17.13 \\
J0958+4222 & 09:58:29.47 & +42:22:56.8 & 2006 Dec 16 &  900 & 2.656 & 18.28 \\
J1009+5450 & 10:09:16.94 & +54:50:03.9 & 2006 Dec 16 &  900 & 2.062 & 19.13 \\
J1019+5246 & 10:19:39.15 & +52:46:27.8 & 2006 Dec 16 &  800 & 2.170 & 17.92 \\
J1029+1039 & 10:29:04.15 & +10:39:01.5 & 2007 Mar 25 &  900 & 1.795 & 17.57 \\
J1049-0110 & 10:49:15.43 & -01:10:38.1 & 2006 Dec 16 &  720 & 2.115 & 17.78 \\
J1054+0348 & 10:54:00.41 & +03:48:01.1 & 2006 Dec 16 &  720 & 2.095 & 17.98 \\
J1056+1208 & 10:56:48.69 & +12:08:26.8 & 2006 Dec 16 &  600 & 1.923 & 17.93 \\
J1057+1506 & 10:57:52.70 & +15:06:14.1 & 2007 Mar 25 & 1000 & 2.169 & 18.06 \\
J1111+1336 & 11:11:19.10 & +13:36:03.8 & 2007 Mar 25 & 1000 & 3.482 & 17.29 \\
J1224+5525 & 12:24:38.42 & +55:25:14.5 & 2007 Mar 25 & 1000 & 1.879 & 17.57 \\
J1310+5424 & 13:10:40.24 & +54:24:49.6 & 2007 Mar 25 & 1000 & 1.929 & 18.50 \\
J1312+5502 & 13:12:01.09 & +55:02:28.6 & 2007 Mar 25 & 1500 & 1.906 & 19.35 \\
J1341+5818 & 13:41:44.63 & +58:18:17.5 & 2007 Mar 25 & 1200 & 2.054 & 18.66 \\
J1357+3450 & 13:57:50.92 & +34:50:23.5 & 2007 Mar 25 & 1400 & 2.921 & 19.03 \\
J1610+4724 & 16:10:09.42 & +47:24:44.5 & 2006 Sep 16 &  900 & 3.217 & 18.76 \\
J1709+3258 & 17:09:09.29 & +32:58:03.4 & 2006 Sep 16 &  600 & 1.889 & 19.22 \\
J2100-0641 & 21:00:25.03 & -06:41:46.0 & 2006 Sep 16 &  900 & 3.130 & 18.17 \\
J2123-0050 & 21:23:29.46 & -00:50:52.9 & 2006 Sep 16 &  120 & 2.262 & 16.45 \\
J2125+0029 & 21:25:21.44 & +00:29:06.4 & 2006 Sep 16 &  900 & 1.951 & 19.30 \\
Q2206-19   & 22:08:51.17 & -19:44:06.5 & 2006 Sep 16 &  300 & 2.559 & 17.0  \\
J2222-0946 & 22:22:56.11 & -09:46:36.2 & 2006 Sep 16 &  600 & 2.926 & 18.00 \\
J2244+1429 & 22:44:52.22 & +14:29:15.1 & 2006 Sep 17 &  700 & 1.955 & 18.94 \\
J2340-0053 & 23:40:23.66 & -00:53:27.0 & 2006 Sep 17 & 1000 & 2.085 & 17.48 \\
\enddata
\tablenotetext{*}{Here we present the list of all observed QSO sightlines.  This list represents four nights of data.  These observations were taken with the MMT BCS using the 800$\frac{grooves}{mm}$ grating.  This gives a wavelength coverage of 3100 to 5100\AA.}
\end{deluxetable*}

The LowRedux pipeline developed by J. Hennawi, S. Burles, and JXP
was used to reduce the MMT data 
\footnote{http://www.ucolick.org/$\sim$xavier/LowRedux}. 
Standard techniques were used to flat-field and bias subtract the raw MMT data.  Objects in the CCD images were traced with a low-order polynomial and extracted optimally.  The resulting 1D spectra were calibrated with a composite HeNeAr and HgCd arc lamp spectrum.  
Instrument flexure was compensated for by shifting the extracted sky spectrum to a 
template spectrum of the sky convolved with the spectral resolution of the BCS.
If more than one spectrum was taken of the same QSO, the multiple exposures 
were coadded into a single spectrum, weighting by the median S/N ratio of each.  
The spectra were fluxed using an observation of a spectrophotometric standard star 
taken on the same night of observation with identical instrumental configuration.
The final data products are wavelength and flux calibrated 1D spectra.   
These are publically available online\footnote{http://www.ucolick.org/$\sim$xavier/DLA/MSDLA}.

For the QSOs that have spectra in the SDSS Data Release 6 
\citep{sdssdr6}, 
we compared the SDSS and MMT spectra visually as a consistency check on our
data reduction.  When comparing the MMT spectra to the SDSS spectra, the relative flux of the MMT spectra did not always match the flux of the SDSS spectra.  We take the SDSS fluxing to be correct and scaled the flux of our MMT spectra correspondingly.  
The correction was a simple multiplicative scalar; we estimate that the relative
fluxing of the MMT is accurate to $\approx 10\%$.
This scaling process is further elaborated upon in our discussion on searching for reddening by MSDLAs in Section \ref{sect:dust}.
  
While comparing the MMT spectrum to the SDSS spectrum from the years 2006.96 and 2002.18 of cMSDLA  J1054+0348, two blueshifted broad absorption line features appeared in the newer MMT spectrum in the that were not present in the older SDSS spectrum.  The time between observations was 4.78 years.  This serendipitous discovery is the first 
observation of a high redshift QSO where associated absorption lines were
observed to appear \citep{hkr+08}.
These absorption lines signify the appearance of a high velocity outflow in the QSO along the line of sight.

No other variable absorption features
were identified in the spectra.

\section{Determination of \nhi\ column densities from DLA \lya\ absorption} \label{sect:nhi}

The principal motivation for the observations analyzed here was to measure the
\ion{H}{1} column densities of $z<2.2$ cMSDLAs.  The MMT observations 
provide spectral coverage of the previously unobserved \lya\ transition.  
With these data, \nhi\ values can be estimated from line-profile fits
to the \lya\ lines \citep[e.g.][]{phw05}.
These \nhi\ values are required to establish the gas metallicity and
to verify that a system satisfies the DLA criterion.  
The cMSDLAs which satisfy both the DLA criterion ($\log \mnhi \ge 20.3$)
and metal-strong criterion (H06) are then classified as MSDLAs.

\begin{figure}
\begin{center}
\includegraphics[width=3.5in]{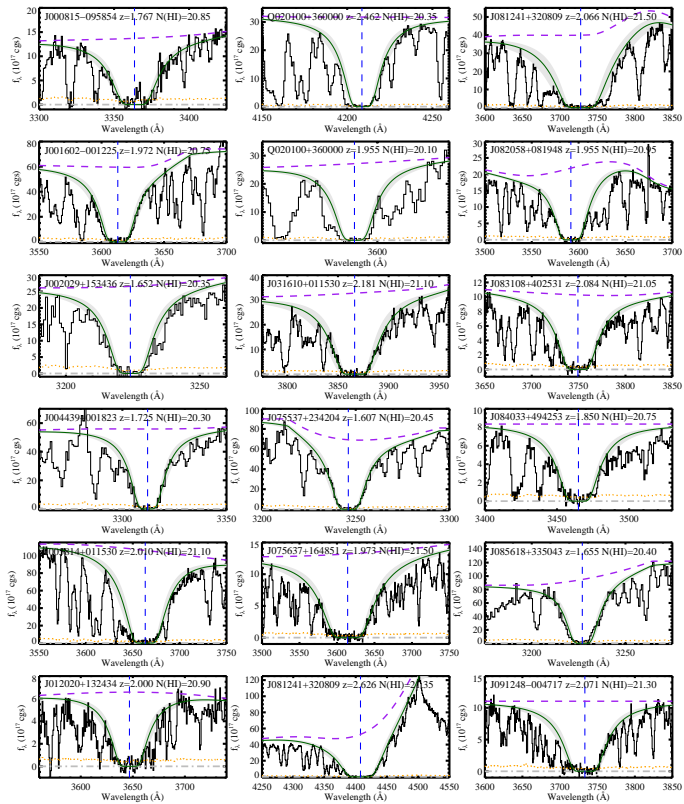}
\end{center}
\caption{
Zoom-in of the \lya\ transition for every strong absorber 
(equivalent width exceeding $\approx 5$\AA) identified
in our MMT spectra.  Overplotted on the data is a model of
the best-fit Voigt profile with the redshift and \nhi\ value
(Table~\ref{tab:hitable})
listed in each panel.  The shaded region indicates an estimate
for the uncertainty, corresponding to $\approx 90\%$ confidence.
The dotted (orange) line traces the 1$\sigma$ error array and the horizontal
dashed (purple) line is our best estimate for the quasar continuum flux.
[See the end of this manuscript for all the fits.]
}
\label{fig:nhifits}
\end{figure}

The \lya\ lines were analyzed using the program x\_fitdla from the XIDL 
package\footnote{http://www.ucolick.org/$\sim$xavier/IDL}.  
For completeness, we considered all \lya\ lines in our dataset
with equivalent widths exceeding $\approx 5$\AA, independent of whether the
system was a cMSDLA.  
For two of the cMSDLAs in our MMT sample (J1111+1336 and J2100$-$0641)
whose redshifts exceed \zabs\ $>3.1$, the \lya\ absorption feature 
was redshifted past our wavelength coverage.
This is because we were searching for molecular hydrogen 
in these QSOs (see Section \ref{sect:metal}).
These DLAs were analyzed using spectra from the SDSS DR6 instead. 
A total of 49 possible DLAs were identified, that includes 
serendipitous DLAs along the sightlines (i.e.\ ones that were
not targeted as MSDLA candidates).  
Redshifts for each DLA were estimated from the centroids of corresponding
low-ion metal transitions, 
with preference given to the \ion{O}{1}~1302 transition as listed in 
Table~\ref{tab:hitable}.

\begin{deluxetable*}{ccccccccc}
\tablewidth{0pc}
\tabletypesize{\footnotesize}
\tablecaption{MMT Sample of DLA Candidates \label{tab:hitable}}
\tablehead{
\colhead{QSO}&\colhead{$z_{abs}$}&\colhead{$log \mnhi^a$}&\colhead{$\log \mnhi^b$}&
\colhead{[M/H]$^c$}&\colhead{$W_{1526}^d$} & \colhead{cMSDLA?}&\colhead{MSDLA?$^e$}\\
&&($\cm{-2})$&($\cm{-2}$)&&(\AA)
}
\startdata
J000815.33-095854 &1.768&$20.85 \pm 0.15$&$20.85 \pm 0.15$&$-0.28 \pm 0.15$&$1.38 \pm 0.10$&Yes & Yes \\
J001602.40-001225 &1.970&$20.75 \pm 0.15$&$20.83 \pm 0.07$&$-0.86 \pm 0.13$&$2.46 \pm 0.10$&Yes & No \\
J002028.96+153436 &1.653&$20.35 \pm 0.15$&$20.35 \pm 0.15$&$-0.51 \pm 0.17$&$0.88 \pm 0.13$&Yes & No \\
J004439.32+001822 &1.725&$20.30 \pm 0.15$&$20.30 \pm 0.15$&$-0.66 \pm 0.12$&$0.89 \pm 0.08$&Yes & No \\
J005814.31+011530 &2.010&$21.10 \pm 0.15$&$21.10 \pm 0.15$&$-0.99 \pm 0.16$&$1.33 \pm 0.05$&Yes & No \\
J012020.37+132433 &1.999&$20.90 \pm 0.15$&$20.90 \pm 0.15$&$-0.56 \pm 0.16$&$0.79 \pm 0.14$&Yes & No \\
Q0201+36 &1.955&$20.10 \pm 0.15$&$20.10 \pm 0.15$&$ $&$...$&No & ... \\
Q0201+36 &2.463&$20.35 \pm 0.15$&$20.38 \pm 0.05$&$-0.36 \pm 0.05$&$...$&Yes & No \\
J031609.83+004043 &2.180&$21.10 \pm 0.20$&$21.10 \pm 0.20$&$-1.20 \pm 0.22$&$0.62 \pm 0.02$&Yes & No \\
J075537.22+234204 &1.670&$20.45 \pm 0.15$&$20.45 \pm 0.15$&$ $&$...$&Yes & ... \\
J075636.73+164850 &1.973&$21.50 \pm 0.15$&$21.50 \pm 0.15$&$ $&$...$&Yes & ... \\
J081240.68+320809 &2.626&$21.35 \pm 0.15$&$21.35 \pm 0.10$&$-0.88 \pm 0.11$&$...$&Yes & Yes \\
J081240.68+320809 &2.067&$21.50 \pm 0.20$&$21.50 \pm 0.20$&$-1.83 \pm 0.20$&$...$&No & No \\
J082058.37+081948 &1.955&$20.95 \pm 0.15$&$20.95 \pm 0.15$&$ $&$1.07 \pm 0.07$&Yes & ... \\
J083108.01+402531 &2.084&$21.05 \pm 0.15$&$21.05 \pm 0.15$&$ $&$1.56 \pm 0.08$&Yes & ... \\
J084032.96+494252 &1.851&$20.75 \pm 0.15$&$20.75 \pm 0.15$&$-0.43 \pm 0.17$&$1.58 \pm 0.13$&Yes & No \\
J085618.23+335042 &1.655&$20.40 \pm 0.15$&$20.40 \pm 0.15$&$ $&$0.62 \pm 0.05$&Yes & ... \\
J091247.59-004717 &2.071&$21.30 \pm 0.15$&$21.30 \pm 0.15$&$-0.89 \pm 0.16$&$...$&Yes & Yes \\
J092708.88+582319 &1.635&$20.40 \pm 0.25$&$20.40 \pm 0.25$&$-0.22 \pm 0.25$&$1.92 \pm 0.15$&Yes & No \\
J093846.77+380549 &1.827&$19.75 \pm 0.15$&$19.75 \pm 0.15$&$ $&$...$&Yes & ... \\
J095829.47+422256 &1.853&$20.90 \pm 0.15$&$20.90 \pm 0.15$&$ $&$...$&Yes & ... \\
J095829.47+422256 &2.065&$20.95 \pm 0.15$&$20.95 \pm 0.15$&$ $&$1.24 \pm 0.06$&Yes & ... \\
J100916.94+545003 &1.623&$20.20 \pm 0.15$&$20.20 \pm 0.15$&$ $&$0.83 \pm 0.13$&Yes & ... \\
J100916.94+545003 &1.892&$21.65 \pm 0.15$&$21.65 \pm 0.15$&$ $&$1.12 \pm 0.11$&Yes & ... \\
J101939.15+524627 &1.833&$19.45 \pm 0.15$&$19.45 \pm 0.15$&$ $&$...$&No & ... \\
J101939.15+524627 &2.018&$20.35 \pm 0.15$&$20.35 \pm 0.15$&$ $&$1.13 \pm 0.05$&Yes & ... \\
J102904.15+103901 &1.622&$21.10 \pm 0.25$&$21.10 \pm 0.25$&$ $&$1.21 \pm 0.07$&Yes & ... \\
J104915.43-011038 &1.658&$20.35 \pm 0.15$&$20.35 \pm 0.15$&$-0.09 \pm 0.15$&$1.74 \pm 0.07$&Yes & No \\
J105400.41+034801 &2.068&$20.40 \pm 0.35$&$20.40 \pm 0.35$&$ $&$1.02 \pm 0.06$&Yes & ... \\
J105648.69+120826 &1.609&$21.45 \pm 0.15$&$21.45 \pm 0.15$&$-0.43 \pm 0.23$&$1.19 \pm 0.09$&Yes & Yes \\
J105752.70+150614 &1.865&$19.35 \pm 0.15$&$19.35 \pm 0.15$&$ $&$...$&No & ... \\
J105752.70+150614 &2.076&$20.50 \pm 0.15$&$20.50 \pm 0.15$&$ $&$1.41 \pm 0.07$&Yes & ... \\
J111119.10+133603 &3.201&$21.25 \pm 0.15$&$21.25 \pm 0.15$&$ $&$...$&Yes & ... \\
J122438.42+552514 &1.673&$20.40 \pm 0.15$&$20.40 \pm 0.15$&$ $&$0.81 \pm 0.04$&Yes & ... \\
J131040.24+542449 &1.600&$20.25 \pm 0.25$&$20.25 \pm 0.25$&$ $&$0.57 \pm 0.16$&No & ... \\
J131040.24+542449 &1.801&$21.45 \pm 0.15$&$21.45 \pm 0.15$&$-0.51 \pm 0.15$&$0.84 \pm 0.12$&Yes & Yes \\
J131201.09+550228 &1.860&$21.20 \pm 0.25$&$21.20 \pm 0.25$&$ $&$2.23 \pm 0.27$&Yes & ... \\
J134144.63+581817 &1.741&$21.00 \pm 0.15$&$21.00 \pm 0.15$&$ $&$1.79 \pm 0.08$&Yes & ... \\
J135750.92+345023 &2.132&$20.45 \pm 0.25$&$20.45 \pm 0.25$&$ $&$...$&Yes & ... \\
J161009.42+472444 &2.508&$21.00 \pm 0.15$&$21.00 \pm 0.15$&$-0.10 \pm 0.15$&$1.76 \pm 0.09$&Yes & Yes \\
J170909.28+325803 &1.830&$20.95 \pm 0.15$&$20.95 \pm 0.15$&$-0.28 \pm 0.15$&$1.62 \pm 0.15$&Yes & Yes \\
J210025.03-064146 &3.092&$21.05 \pm 0.15$&$21.05 \pm 0.15$&$-0.67 \pm 0.15$&$1.09 \pm 0.04$&Yes & No \\
J212329.46-005052 &2.058&$19.45 \pm 0.15$&$19.45 \pm 0.15$&$ $&$0.48 \pm 0.03$&Yes & ... \\
J212521.44+002906 &1.751&$21.35 \pm 0.20$&$21.35 \pm 0.20$&$ $&$1.71 \pm 0.23$&Yes & ... \\
Q2206-19 &1.920&$20.65 \pm 0.15$&$20.65 \pm 0.07$&$-0.37 \pm 0.07$&$0.99 \pm 0.01$&Yes & No \\
Q2206-19 &2.076&$20.50 \pm 0.15$&$20.43 \pm 0.06$&$-2.26 \pm 0.07$&$...$&No & No \\
J222256.11-094636 &2.354&$20.55 \pm 0.15$&$20.55 \pm 0.15$&$-0.61 \pm 0.16$&$1.20 \pm 0.06$&Yes & No \\
J224452.22+142915 &1.816&$20.70 \pm 0.15$&$20.70 \pm 0.15$&$-0.59 \pm 0.16$&$1.08 \pm 0.13$&Yes & No \\
J234023.66-005327 &2.055&$20.35 \pm 0.15$&$20.35 \pm 0.15$&$-0.59 \pm 0.15$&$...$&Yes & No \\
\enddata
\tablenotetext{a}{Measured from our MMT observations.}
\tablenotetext{b}{Adopted value based on comparison of our measurement with ones in the literature (see text).}
\tablenotetext{c}{{L}ogarithmic metallicity relative to solar as measured from ionic transitions of Zn$^+$, Si$^+$, or S$^+$.  The measurements adopt the solar abundances of \cite{asplund09}.}
\tablenotetext{d}{Rest-frame equivalent width of the \ion{Si}{2}~1526 transiiton.  In nearly every case, this was measured from a Gaussian fit to the observed line in the SDSS spectrum.  The remainder are boxcar integrations of higher resolution observations.}
\tablenotetext{e}{DLAs without an entry have not been observed at sufficiently high spectral resolution to test the metal-strong criteria.  Entries that are cMSDLA and have $\log \mnhi \ge 20.3$ but are not MSDLAs have metal column densities below the metal-strong criteria.}
\end{deluxetable*}

With the redshift of the possible DLA constrained, we measured the \ion{H}{1} column density
as follows.  First, the continuum from the background QSO is modeled by hand in the
region local to the \lya\ line.  The centroid of the \lya\ line is given by the previously measured redshift.
Then a Voigt profile is fit to the \lya\ absorption feature by visual inspection.  The value of \ncol\ is adjusted until the wings of the profile best fit the observed \lya\ line.
The continuum and \nhi\ values were then iteratively modified (generally by $<10\%$
from our initial guess) until an optimal match to the data was established.
The resulting profile gives the \ion{H}{1} column density directly.  All of
the \nhi\ values are listed in Table~\ref{tab:hitable}. 
The errors in these values are dominated by systematic uncertainties, primarily
continuum placement and line-blending in the damped wings of the Voigt profile.
In all cases, we assumed a minimum error of 0.15\,dex.  
We estimate larger uncertainties for DLAs
where the QSO continuum is especially challenging, there is significant
emission-line contamination from the QSO and/or if there is substantial 
line-blending with the \lya\ forest.  All of
our Voigt profile fits are presented in Figure \ref{fig:nhifits}.

%\begin{figure} [p]
%\plotone{f1b.eps}
%\includegraphics[width=\textwidth, height=\textheight] {../Figures/fig_fits1.ps}
%\end{figure}
%\begin{figure} [p]
%\plotone{f1c.eps}
%\includegraphics[width=\textwidth, height=\textheight] {../Figures/fig_fits2.ps}
%\end{figure}

\begin{figure}
\centering
\begin{center}
\includegraphics[height=3.5in,angle=90]{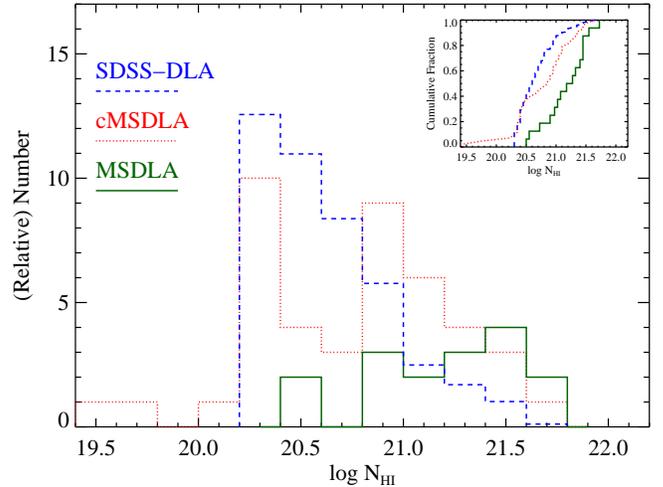}
\end{center}
\caption{
Histogram comparing the \ion{H}{1} column densities of the 
MSDLA candidates (cMSDLA; red, dotted), 
confirmed MSDLAs (MSDLA; green, solid), 
and a comparison sample (blue, dashed)
of $z<3$ DLAs from Data Release 5
of the Sloan Digital Sky Survey \citep[SDSS-DLA][]{pw09}.  
The SDSS histogram (scaled to have identical area under the
curve as the cMSDLA histogram) represents the distribution of
\nhi\ values for randomly selected quasar sightlines at $z \sim 3$
and serves as a control sample.
A normalized cumulative plot is shown in the top right corner.
The cMSDLA and especially the confirmed MSDLA exhibit significantly
higher \nhi\ value on average.  A Kolmogorov-Smirnov test comparing the cMSDLA
and MSDLA against the SDSS control sample gives KS probabilities of
$P_{KS} = 0.002$ and $10^{-5}$ respectively. 
}
\label{fig:nhihist}
\end{figure}

We have compared our values with existing estimates from the literature.
Previous measurements of H I column densities were found for eight
of the absorption systems we observed.    Four of these previous measurements from the literature were made at significantly higher spectral resolution \citep{pw96, p03_esi, pro01}.  We have adopted these H I column density measurements over our own (see fourth column of Table \ref{tab:hitable}).  In all cases the difference is well within our estimated error.
We caution, however, that because the uncertainties are dominated by systematic
error that are largely independent of the spectral resolution and signal-to-noise,
the good agreement does not mean that we have overestimated the uncertainties
\citep[see also][]{phw05}.

We targeted 43 systems as cMSDLAs, most of which had no prior H I column density measurements.
Beyond the 43 absorption systems targeted, five QSO slight lines contain a second absorption system.  This brings the total amount of \lya\ absorbers to 49.
Out of the 49 \lya\ absorption systems identified, 42 have \ion{H}{1} column 
densities that satisfy the DLA criterion (\ncol $\ge 20.3 $ \atomspercm). 
The remainder are super Lyman limit systems (SLLS) with $\log \mnhi \ge 19$. 
%Of the full list, 43 were targeted as cMSDLAs.
All but two of the cMSDLAs are confirmed as 
damped \lya\ systems.
The two SLLS that were also cMSDLAs 
(J2123$-$0050 at $z_{\rm abs} = 2.058$ and J1009+5450 at $z_{\rm abs}=1.623$) 
may represent new examples of super-solar Lyman limit
systems \citep[e.g.][]{peroux06,pho+06}.

Figure \ref{fig:nhihist} shows a histogram of H I column densities 
for MSDLAs, our sample of cMSDLAs, and a comparison sample of randomly 
selected DLAs from the SDSS-DR5 \citep{pw09}.  
The latter sample contains 380 DLAs with 
$z_{\rm abs} = [2.2,3.0]$.  Although the 
median redshift of this control sample is higher than
metal-strong systems, \cite{pw09} have demonstrated that 
evolution in the shape of the \ion{H}{1} frequency distribution 
is weak for DLAs.
The MSDLAs and cMSDLAs show systematically higher \ion{H}{1} column densities then 
the comparison sample.  
Formally, a two-sided KS test yields probabilities $P_{\rm KS} = 0.002, 10^{-5}$ 
that cMSDLAs and MSDLAs respectively are drawn from the
same parent population as the control sample.
This is consistent with the expectation that selecting 
systems with high metal column densities will select systems with high 
\ion{H}{1} column densities.

\section{Metal Column Density and Metallicity Measurements} \label{sect:metal}

To estimate the metallicities of the MSDLAs we simply compare the measured \ion{H}{1}
column density with the ionic column density of an element expected to trace the
metal abundance of the system.  Generally, one avoids refractory elements like Fe
because these can be depleted from the gas-phase and may give systematically low
metallicity values.   Previous work has focused on low-ion transitions of Si, S, and
Zn which are mildly refractory, give similar results, and have several low-ion 
transitions for analysis \citep[e.g.][]{pw01,pgw+03}.
Unfortunately, the MMT and SDSS spectra have insufficient spectral
resolution for a precise column density measurement.  Therefore,
we have adopted the ionic column densities for these DLAs from previous works
\citep[H06;][]{pwh+07} and new Keck/HIRES observations that will be presented in a
future paper (Prochaska et al., in prep.).  Measurements of metal column densities exist for 24 of the absorbers in our sample, 22 of which are cMSDLAs.
These metal column density measurements test whether a cMSDLA
satisfies the metal-strong criteria set by H06:
$\log \N{Zn^+} \ge 13.15$ and/or $\log \N{Si^+} \ge 15.95$.
Those cMSDLAs that meet the metal-strong criteria are classified as MSDLAs.
Out of 22 cMSDLAs with metal column density measurements, 7 meet the metal-strong criteria and are classified as MSDLAs.
When combined with measurements from the literature, we have
a total sample of 15 MSDLAs with $z_{\rm abs} \ge 1.60$.

\begin{figure}[ht]
\begin{center}
\includegraphics[height=3.5in,angle=90]{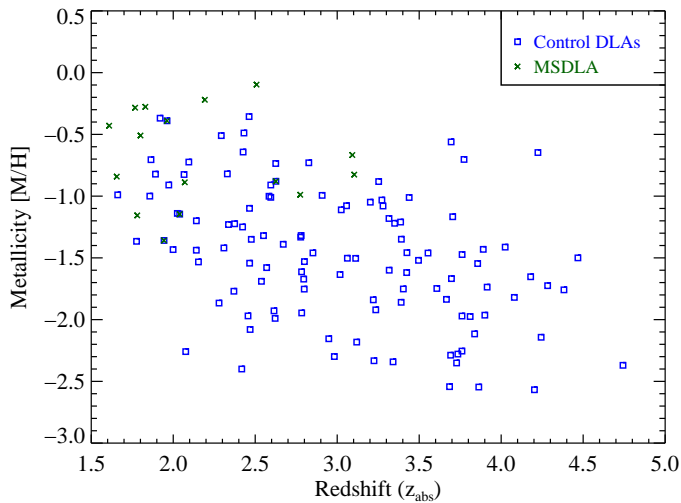}
\end{center}
\caption{Scatter plot of gas metallicity 
([M/H]; determined from Si, S, O, or Zn) against absorber redshift.
The blue squares show the metallicity values for a (nearly) random 
DLA sample \citep{pgw+03} and the (green) x's show values for
the MSDLA systems.
We find that the MSDLA systems exhibit
systematically higher metallicity than the random sample to at least
$z \approx 3$.
}
\label{fig:zvsmtl}
\end{figure}

The metallicity of a DLA is usually expressed as the logarithmic ratio of the 
metal to \ion{H}{1} column densities relative to the solar abundance, [M/H].  
Here, the letter `M' refers to metallicity, derived from one of the
elements listed above.  For the solar abundance scale, we adopt
the compilation of \citep{asplund09} using meteoritic values when available
excepting C, N, and O. 
The values are listed in Table~\ref{tab:hitable}.  
These new metallicity values allow investigation of several relationships.
In the universe, metallicity increases over time from the original 
abundances created in Big Bang nucleosynthesis as stars evolve and die 
enriching their surrounding interstellar media.  One would expect that 
metallicities of DLAs to decrease with 
greater lookback time (here quantified by increasing absorber redshift).  
In Figure \ref{fig:zvsmtl}, the
metallicity of the DLAs is compared with the absorber's redshift.  The 
DLAs from \cite{pgw+03}, which are taken as a random sample,
follow the expected trend.  The MSDLAs exhibit 
systematically higher metallicity than the random sample over the range of 
redshifts covered.  
Specifically, we measure a median metallicity for the MSDLA of
[M/H]~$= -0.67$\,dex over the redshift interval $z_{\rm abs}= [1.6,3.1]$
and calculate a median [M/H]~$= -1.32$\,dex from the control sample over the same interval.
We expect this result since the MSDLAs are selected for their high metal column densities,
which should select systems with high metallicities across a range of redshifts.

%It is worth noting that the MSDLAs, because of their method
%of discovery, tend to have $z_{\rm abs}$ closer to $z_{\rm em}$
%than most other DLA surveys.  Of our sample, two lie within 5000\kms\
%of the quasar.

DLAs, by definition, contain a large amount of neutral hydrogen.  It can be expected that some might also contain large amounts of molecular hydrogen (H$_2$), especially if the QSO sight-line probes a high density region of the DLA. H$_2$ has been detected in DLAs by previous authors \citep[e.g.][]{gb97,petit00}.  Three of 
the cMSDLAs show evidence of H$_2$ in their spectra.
These systems are J0812+3208 ($z_{\rm abs} = 2.626$), 
J2100$-$0641 ($z_{\rm abs}=3.092$), and J2340$-$0053 ($z_{\rm abs} = 2.054$).
The remainder of the systems show no obvious H$_2$ absorption at the
strong transitions of the $J=0$ and $J=1$ levels 
(i.e.\ $N(\rm H_2) < 10^{17}$ \atomspercm).  We conclude that
only 3 out of the 32 systems have significant $H_2$ absoprtion giving
a detection rate of $\approx 10\%$. 
This is a surprisingly low detection rate given that the systems
have systematically higher metallicity than random samples of DLAs.
Indeed, \cite{nlp+08} report a detection rate of $\approx 35\%$
for systems with [M/H]~$\ge -1.3$ to a similar limit on $N(\rm H_2)$.
The incidence is higher
(2 of 6 for systems with good coverage of the Lyman-Werner transitions) 
if we restrict the discussion to MSDLAs, but this sample is
too small for statstical discussion.

\begin{figure}[ht]
\begin{center}
\includegraphics[height=3.5in,angle=90]{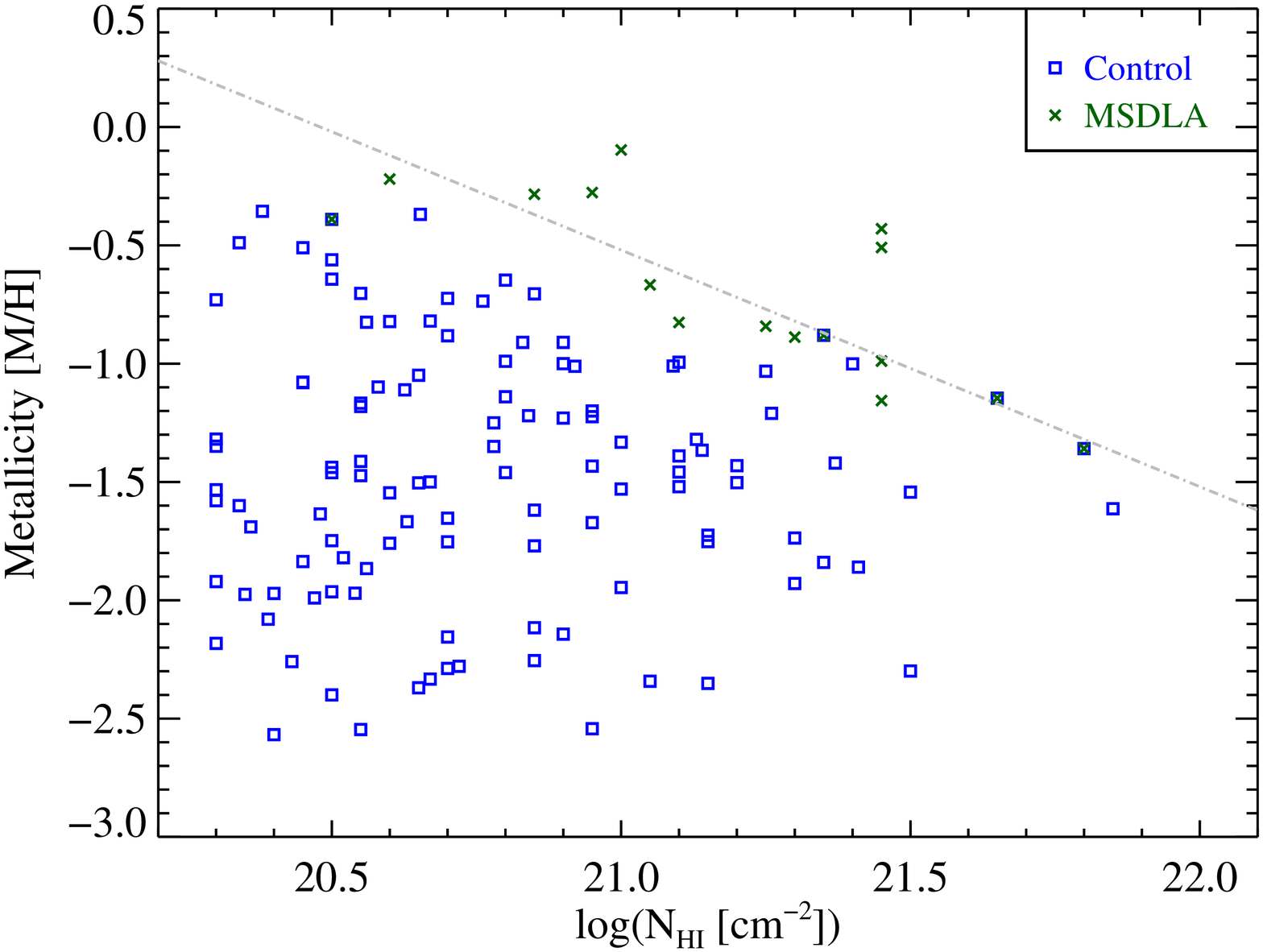}
\end{center}
%\centering
%\includegraphics[height=\textwidth, angle=90]{../Figures/nhi_vs_mtl.ps}
\caption{Scatter plot of gas metallicity 
([M/H]; determined from Si, S, O, or Zn) against absorber \nhi\ value,
restricted to those exceeding the damped \lya\ criterion.
The blue squares show the metallicity values for a (nearly) randomly selected
DLA sample \citep{pgw+03} and (green) x's show values for
the MSDLA systems.
The overplotted dash-dot line traces the `upper-bound' to the [M/H], \nhi\
distribution analyzed by \cite{blb+98}.
The MSDLA, by definition, follow or exceed this supposed boundary.
Our results indicate that dust obscuration does not entirely
preclude the detection
of sightlines with very high metal column density.
}
\label{fig:nhivsmtl}
\end{figure}

Figure~\ref{fig:nhivsmtl}
compares the metallicity of DLAs to their \ion{H}{1} 
column densities (\nhi)  for the MSDLAs and cMSDLAs along with 
the same random sample \citep{pgw+03}. 
%Commented out at the suggestion of Mirka
%MSDLAs again exhibit systematically higher metallicities then the random sample.  
Most of our confirmed MSDLAs lie on or beyond the upper bound on metallicity/\nhi\ discussed by \cite{blb+98}, who suggested this upper bound is caused by obscuration of the background QSO by large amounts of dust.  In fact, MSDLAs must by definition be near or above this upper bound.  Our observations of MSDLAs clearly lie beyond this limit, 
indicating that dust does not always completely obscure systems with 
very high metal column densities (see also H06).  
This lends further support to prior assertions that this observational bound
was not driven entirely by obscuration \citep{ehl05,kep+09}.
The consistently higher metallicity of MSDLAs could incidate that their 
sightlines probe regions near the center of galaxies (or proto-galaxies) 
if a radial gradient in metallicity exists
for high $z$ galaxies.  Alternatively, the higher metallicity may indicate
that MSDLAs probe more chemically evolved systems and, if a mass-metallicity
relation holds at high $z$ \cite[e.g.][]{esp+06,pcw+08}, higher mass galaxies.

\begin{figure}
\begin{center}
\includegraphics[height=3.5in,angle=90]{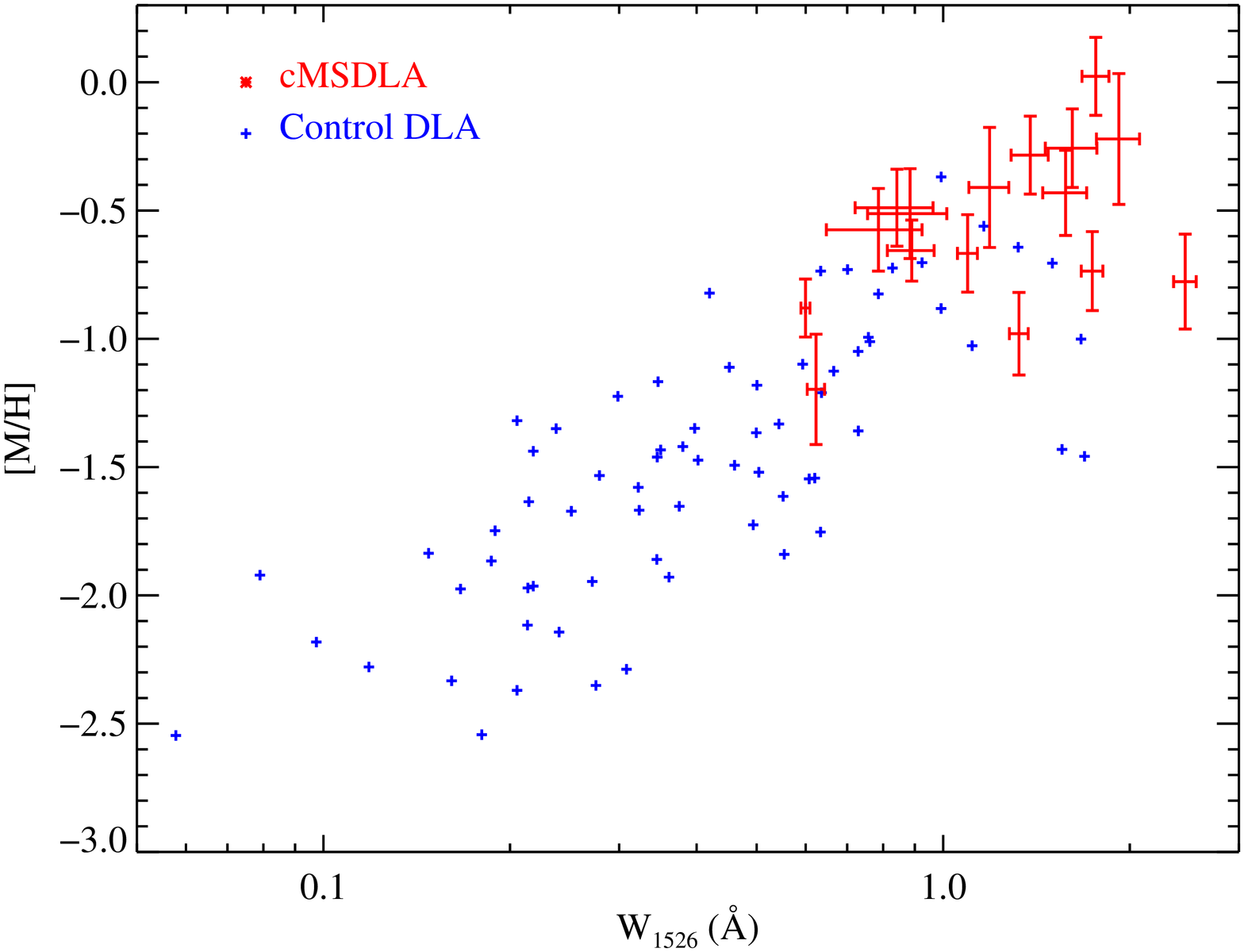}
\end{center}
\caption{Gas metallicity values versus rest-frame equivalent width
of the \ion{Si}{2}~1526 transition ($W_{1526}$) as measured for the random sample
of DLAs from \cite{pgw+03} and our MSDLA candidates.  
Our new observations indicate that the [M/H]/$W_{1526}$ correlation 
holds to nearly solar metallicity and very large $W_{1526}$ values.
If interpreted as a mass-metallicity relation \citep{pcw+08},
this implies the MSDLAs represent the most massive galaxies of the DLA
population.  Errors for the contorl DLAs are negligible for \siew\ and
are $0.1-0.15$\,dex for the [M/H] values.
}
\label{fig:si2ew}
\end{figure}

To test the latter hypothesis, we examined the relation
between \ion{Si}{2}~1526 equivalent width \siew\ and the gas metallicity.
A random sample of DLAs was shown by
\cite{pcw+08} to exhibit a correlation between these two measures,
which the authors interpret as a mass-metallicity relation
\citep[see also][]{mcw+07}.  
The key point is that \siew\ is dominated by the kinematics
of the gas, generally weaker `clouds' that contribute little to the column
density of Si$^+$. 
More massive systems have higher gravitational potentials, thus faster 
rotational and/or virial motion which leads to a greater 
broadening of \ion{Si}{2}~1526.     
\siew\ values for the systems in this sample have 
been measured using Keck HIRES and ESI spectra by 
\cite[H06;][]{pwh+07} and Kaplan et al. (in prep).  These metal-strong systems 
probe a higher range of metallicities than previous observations, extending 
the \w\ vs.  [M/H] relation.  
Figure \ref{fig:si2ew} shows this relationship for our data 
and a random sample of DLAs with lower metallicities taken from \cite{pgw+03}.  

Our new measurements indicate
the \siew-metallicity relationship holds to solar metallicity with roughly
constant scatter.  If there is an underlying mass-metallicity relationship 
then the MSDLAs, with their systematically higher metallicities, also represent the
subset of DLAs with highest mass.
These are ideal candidates for follow-up imaging surveys if one wishes
to maximize observing efficiency \citep{mff04}.

\section{Dust Extinction} \label{sect:dust}

Dust in the Galaxy's interstellar medium is known to absorb and scatter the light from background objects.   
Dust consists of grains of heavy elements that have 
solidified out of the atmospheres of cool stars and the interstellar medium.  
MSDLAs, with their high metal column densities, make good candidates to 
search for dust in the high redshift universe and set (rare)
constraints on the properties of such dust, e.g.\ the dust-to-metals ratio,
extinction laws.

%Dust absorbs and scatters incident light from background objects.  
This absorption is most efficient at short wavelengths, 
and has the tendency to redden the color of the background light source.  
This process is known as extinction and approximately follows a $\lambda^{-1}$ relation.  
In the case of QSO absorption systems like DLAs, dust in the DLA reddens the color 
of the background QSO. 
Extinction ($A_\lambda$) is defined as the change in magnitude due to dust absorption.  
Reddening is the difference in extinction  
between two bands, most commonly reported between $B$ and $V$
as $E_{B-V} \equiv A_B-A_V$.  

Extinction in a large sample of DLAs from the SDSS have been 
studied photometrically.   \cite{vpw08} found evidence for reddening 
in a large sample of DLAs when compared to QSOs that contain no absorbers.  
Previously, \cite{vcl+06} found evidence for reddening 
in 5 out of 13 DLAs selected to have strong \ion{Zn}{2} absorption 
lines.  Several of these 5 systems qualify as MSDLAs.  In fact  
two are also in our sample of MSDLA candidates 
(J0016$-$0012 and J2340$-$0053) and both show evidence of reddening.   
The high metal column densities in MSDLAs suggest that they may contain 
a large amount of dust.  If photometric studies of DLAs find evidence for 
dust in two systems that are metal strong candidates, it is plausible that 
many more cMSDLAs will also yield evidence for dust.

\begin{figure*}
\begin{center}
\includegraphics[height=6.8in,angle=90]{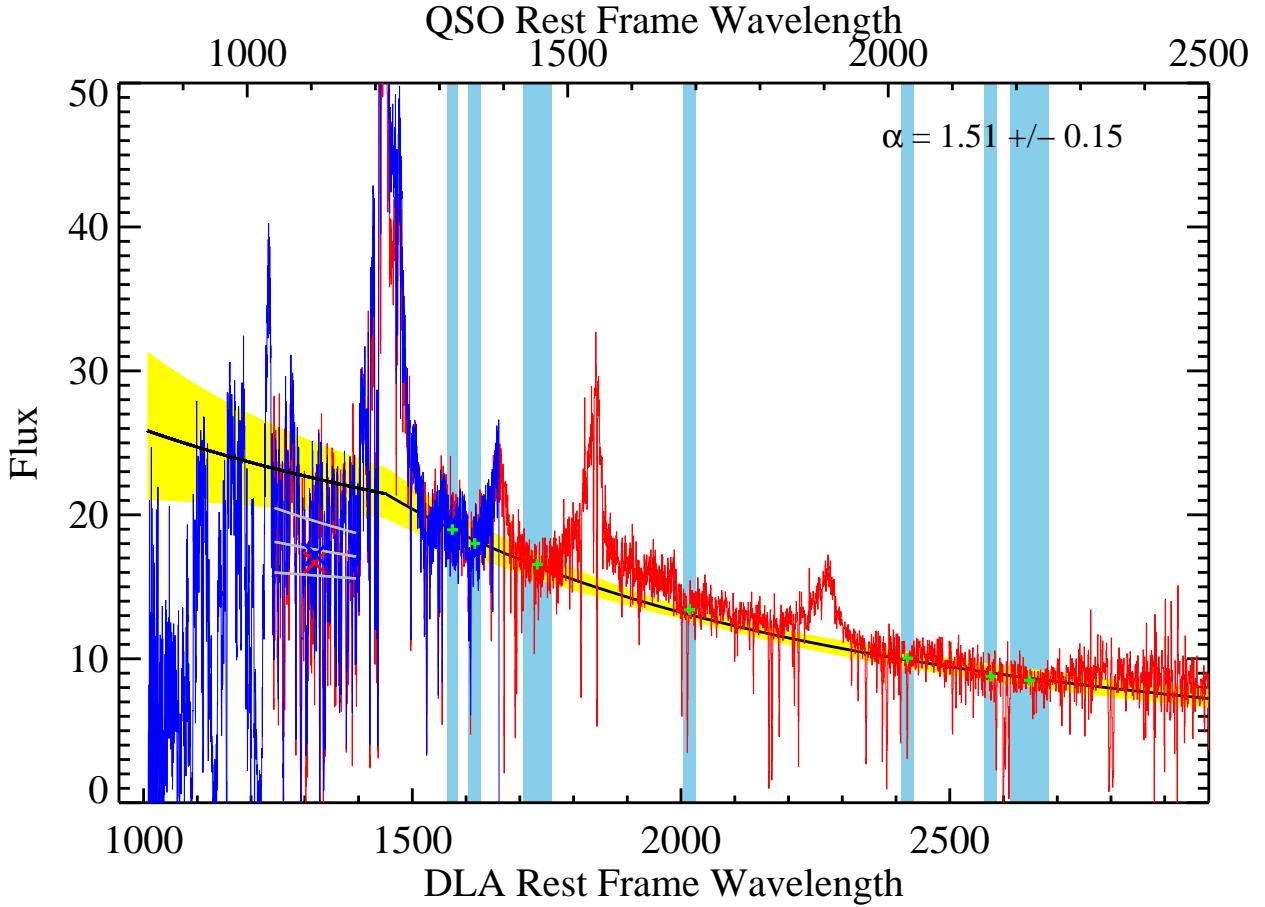}
\end{center}
%\centering
%\includegraphics[height=\textwidth, angle=90]{../Figures/fig_powerlaw.ps}
\caption{SDSS (red; $\lambda_{QSO} > 800$\AA) and 
MMT (dark blue; $\lambda_{QSO} < 1200$\AA) spectra of the quasar 
J0958+4222 ($z_{em} = $2.656).  The upper x-axis shows the wavelengths
in the quasar rest-frame while the lower x-axis refers to the
rest-frame of the foreground candidate metal-strong damped \lya\ system (cMSDLA;
$z_{abs} = $2.065).  The vertical shaded regions indicate 
`emission-free' windows in the quasar continuum, which were used
to model a power-law spectrum $f_\lambda \propto \lambda^{-\alpha}$
to the SDSS data.
The (green) plus signs show the median flux in each region.
The black solid line shows the best-fit model, normalized at 
$\lambda_{QSO} = 1700$\AA\ and the yellow shaded region about the
line indicates the $1\sigma$ uncertainty.  A second power-law, with an exponent that differs by $\Delta\alpha=1.0$,
 describes the model quasar continuum blueward of $\lambda_{QSO} = 1216$\AA.
%The black horizontal line at $\lambda_{QSO} \approx 1000$\AA\ indicates where the DLA's
%\lya\ absorption line is masked out.
The short, grey lines at $\lambda_{QSO} \approx 1100$\AA\ 
show the predicted mean flux (and the uncertainty)
of the continuum when accounting for 
absorption by the \lya\ forest \citep{kts+05}.
The X's show the mean flux for the MMT (blue) and SDSS (red) spectra in the region indicated by the grey lines.
}
\label{fig:powerlaw}
\end{figure*}

\subsection{Quasar Power-Law Slopes} \label{subsec:powerlawslope}

In principle, analysis of 
spectrophotometric observations for reddening are ideal because the extinction 
can be explored as a function of wavelength.
In practice, this is difficult because the intrinsic shape 
of the background QSO continuum is unknown and is likely to vary
with significant scatter.  
%JXP -- The same is true for photometry.  Spectra actually helps because
%  you can explicitly avoid emission lines.
While this makes constraining reddening for individual systems difficult, 
one may explore systematic shifts in the overall population by
essentially averaging over the scatter in intrinsic QSO continua. 
A spectroscopic search for reddening in DLAs was previously 
conducted by \cite{ml04} who fit a power-law ($f_\nu \varpropto \nu^\alpha$) to the QSO continuum for each system and compared the exponent $\alpha$ to a control sample of QSOs.  These DLAs were not selected to be metal strong, and  
the authors did not detect reddening along the 70 DLA sightlines.  
We now conduct the same type of analysis, but with our sample of cMSDLAs.

The typical QSO spectrum in the optical and ultraviolet consists of two distinct features.  The continuum spans across all wavelengths and is shaped by one or 
more power-laws \citep[e.g.][]{telfer02}.  Broad emission lines, such as \lya, lie on top of the continuum.  Our spectra of cMSDLAs taken with the MMT/BCS covers a wavelength range of $\sim 3100$ to 5100\AA, the blue end of the visible spectrum.  Most of the cMSDLAs also have corresponding SDSS DR6 spectra, covering a wavelength range of $\sim$ 3800 to 9200 \AA.  Both the MMT and SDSS spectra provide a wide coverage of wavelengths spanning the UV spectrum in the QSO rest-frame.

The first step in this analysis is to model the QSO's continuum spectrum by fitting a power-law to the regions in the spectrum free from QSO emission lines:
\begin{equation} \label{eqtn:powerlaw}
f_r(\lambda) = f_{obs}(\rm 1700 \AA) \ltp \frac{\lambda}{\rm 1700 \AA} \rtp^{-\alpha}.
\end{equation}
See Figure~\ref{fig:powerlaw} for an example of our power-law model.  
This is based on a similar method created by \cite{ml04}.
The seven `emission free' (EM-free) regions are 
1312-1328,  1345-1365, 1430-1475, 1680-1700, 2020-2040, 2150-2170, and 2190-2250 \AA\ in the QSO rest frame.  We mask emission-free regions which are unreliable or contaminated by absorption lines from the DLA.
The median flux and central wavelength in each emission free region form 
points from which the power-law curve is fitted.  We normalize our power-law fit to the QSO's flux at 1700 \AA.  This fit is done in the SDSS spectrum because the fluxing is presumed to be accurate and the SDSS spectrum covers longer wavelengths redward of the QSO's \lya\ line where the \lya\ forest cannot 
contaminate the continuum flux.
We derived a statistical error on $\alpha$ using the IDL program 'curvefit.'  These ranged from from approximately $1-50\%$ in $\alpha$.  It is evident from inspection of the data and model that these are frequently poor estimates of the total uncertainty.  We estimate the total uncertainty (statistical and systematic) to be $15-20\%$.

\begin{figure}
\begin{center}
\includegraphics[height=3.5in,angle=90]{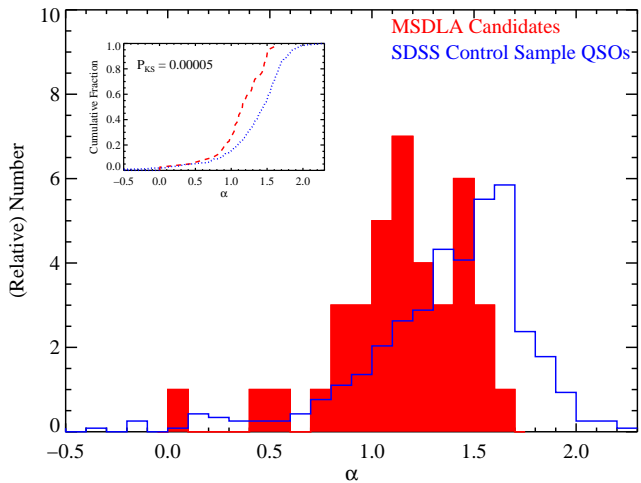}
\end{center}
\caption{
Histogram of power-law exponents ($\alpha$) for the MSDLA candidates
(red) compared against a control sample from the SDSS-DR5
taken to have the same $r$-magnitude and redshift distribution
as the MSDLAs (normalized to have the same area under the curve).
Larger $\alpha$ values indicate a bluer quasar spectrum.
The upper left plot shows the normalized, cumulative
distribution.  The MSDLA candidates show systematically lower
$\alpha$ values indicating reddening by dust within the galaxies.
The difference in median $\alpha$ between the control sample and cMSDLAs
is $\delta \alpha = 0.29$.
}
\label{fig:histalpha}
\end{figure}

The exponent of the power-law ($\alpha$) presumably becomes depressed when the background QSO is reddened by dust.  The analysis by \cite{ml04} could not find any significant difference between the exponents for ordinary 
non-metal strong DLAs and a control sample of QSOs.  The same comparison is done using our sample of cMSDLAs, as seen in Figure \ref{fig:histalpha}.  
(Table~\ref{tab:dust}).
For each cMSDLA,
we construct a control sample of $\sim12$ randomly selected QSOs 
with a similar redshift ($\Delta$\zabs=0.2) and $r$-magnitude ($\Delta r$=0.355\,mag).
The latter value was chosen to give at least 12 systems in the control
sample for each cMSDLA.
The cMSDLA sample exhibits systematically lower $\alpha$ then the 
combined control sample, 
indicating they are reddening their background quasars.
This is true for both median (see below) and mean statistics indicating
there is reddening by the bulk of the cMSDLAs.
This contrasts with the results from \cite{ml04} who found no 
conclusive evidence for reddening in `random' DLAs.  
The difference in the median exponent of the control sample $\alpha_C$
and the median exponent of the cMSDLAs $\alpha_m$
is defined to be $\delta\alpha \equiv \alpha_C-\alpha_m$.  
We measure $\alpha_C=1.45$ and $\alpha_m= 1.16$ giving $\delta\alpha=0.29$.  
In the following section we use $\delta\alpha$ 
to constrain the typical extinction and reddening by dust in these systems.

%    MSDLAs, with their relatively high dust content, must represent a unique subset of normally dust poor high redshift DLAs.

\begin{deluxetable*}{cccccccc}
\tablewidth{0pc}
\tabletypesize{\footnotesize}
\tablecaption{Calculated alpha and dust absorption parameters \label{tab:dust}}
\tablehead{
\colhead{QSO}&
\colhead{$z_{em}$}&
\colhead{$z_{abs}$}&
\colhead{$\alpha_r$}&
\colhead{$\alpha_b$}&
\colhead{Dust Absorption}&
\colhead{Dust Absorption}&
\colhead{$\log \mneff^a$}
\\ &&&&& (MMT) & (SDSS) & (cm$^{-2}$)
}
\startdata
J0008-0958 & 1.9512 & 1.7669 & 1.27 & 0.27 &$0.430\pm0.490$&$...$& 20.57 \\
J0016-0012 & 2.0869 & 1.9720 & 1.05 & 0.05 &$-0.011\pm0.055$&$...$& 19.97 \\
J0020+1534 & 1.7635 & 1.6521 & 1.16 & 0.16 &$...$&$...$& 19.84 \\
J0044+0018 & 1.8677 & 1.7245 & 1.29 & 0.29 &$0.067\pm0.141$&$...$& 19.64 \\
J0058+0115 & 2.4949 & 2.0134 & 1.60 & 0.60 &$0.039\pm0.110$&$0.034\pm0.105$& 20.12 \\
J0120+1324 & 2.5671 & 1.9998 & 1.50 & 0.50 &$-0.007\pm0.227$&$0.076\pm0.286$& 20.32 \\
J0316+0040 & 2.9206 & 2.1805 & 0.98 & -0.02 &$-0.027\pm0.152$&$-0.037\pm0.145$& 19.90 \\
J0755+2342 & 1.8249 & 1.6702 & 1.29 & 0.29 &$0.062\pm0.109$&$...$& ... \\
J0756+1648 & 2.8656 & 1.9730 & 1.05 & 0.05 &$-0.140\pm0.124$&$-0.209\pm0.070$& ... \\
J0812+3208 & 2.7105 & 2.6257 & 1.02 & 0.02 &$0.076\pm0.129$&$0.056\pm0.110$& 20.47 \\
J0820+0819 & 2.5191 & 1.9545 & 1.22 & 0.22 &$-0.101\pm0.049$&$-0.081\pm0.060$& ... \\
J0831+4025 & 2.3300 & 2.0841 & 1.33 & 0.33 &$0.063\pm0.225$&$0.117\pm0.254$& ... \\
J0840+4942 & 2.0763 & 1.8502 & 0.97 & -0.03 &$0.060\pm0.236$&$...$& 20.32 \\
J0856+3350 & 1.7256 & 1.6545 & 1.16 & 0.16 &$-0.193\pm-0.136$&$...$& ... \\
J0912-0047 & 2.8593 & 2.0710 & 1.49 & 0.49 &$-0.062\pm0.153$&$-0.095\pm0.128$& 20.43 \\
J0927+5823 & 1.9100 & 1.6350 & 0.01 & -0.99 &$-0.044\pm0.011$&$...$& 20.18 \\
J0938+3805 & 1.8274 & 1.8266 & 0.58 & -0.42 &$-0.205\pm-0.147$&$...$& ... \\
J0958+4222 & 2.6558 & 2.0650 & 1.51 & 0.51 &$0.022\pm0.120$&$0.053\pm0.148$& ... \\
J1009+5450 & 2.0616 & 1.8922 & 1.09 & 0.09 &$0.172\pm0.358$&$...$& ... \\
J1019+5246 & 2.1701 & 2.0181 & 1.43 & 0.43 &$0.094\pm0.152$&$...$& ... \\
J1029+1039 & 1.7946 & 1.6221 & 1.04 & 0.04 &$0.135\pm0.216$&$...$& ... \\
J1049-0110 & 2.1153 & 1.6580 & 1.13 & 0.13 &$-0.092\pm0.012$&$...$& 19.61 \\
J1054+0348 & 2.0947 & 2.0682 & 1.45 & 0.45 &$0.087\pm0.144$&$...$& ... \\
J1056+1208 & 1.9227 & 1.6091 & 1.20 & 0.20 &$0.343\pm0.420$&$...$& 21.04 \\
J1057+1506 & 2.1690 & 2.0761 & 1.64 & 0.64 &$0.100\pm0.158$&$...$& ... \\
J1111+1336 & 3.4816 & 3.2011 & 1.46 & 0.46 &$-0.141\pm-0.043$&$-0.162\pm-0.065$& ... \\
J1224+5525 & 1.8788 & 1.6727 & 1.45 & 0.45 &$0.118\pm0.165$&$...$& ... \\
J1310+5424 & 1.9292 & 1.8004 & 0.43 & -0.57 &$0.061\pm0.151$&$...$& 20.96 \\
J1312+5502 & 1.9065 & 1.8604 & 0.91 & -0.09 &$0.362\pm0.533$&$...$& ... \\
J1341+5818 & 2.0542 & 1.7407 & 1.10 & 0.10 &$0.111\pm0.191$&$...$& ... \\
J1357+3450 & 2.9214 & 2.1316 & 0.86 & -0.14 &$-0.059\pm0.207$&$-0.068\pm0.201$& ... \\
J1610+4724 & 3.2169 & 2.5078 & 1.15 & 0.15 &$0.201\pm0.448$&$0.095\pm0.375$& 21.02 \\
J1709+3258 & 1.8891 & 1.8300 & 0.82 & -0.18 &$0.062\pm0.259$&$...$& 20.69 \\
J2100-0641 & 3.1376 & 3.0920 & 1.50 & 0.50 &$0.128\pm0.240$&$0.052\pm0.173$& 20.38 \\
J2123-0050 & 2.2623 & 2.0582 & 1.39 & 0.39 &$-0.028\pm0.022$&$0.022\pm0.065$& ... \\
J2125+0029 & 1.9505 & 1.7510 & 1.11 & 0.11 &$0.457\pm0.563$&$...$& ... \\
J2222-0946 & 2.9263 & 2.3545 & 0.89 & -0.11 &$-0.306\pm-0.204$&$-0.294\pm-0.193$& 19.94 \\
J2244+1429 & 1.9546 & 1.8153 & 1.32 & 0.32 &$0.129\pm0.284$&$...$& 19.81 \\
J2340-0053 & 2.0845 & 2.0541 & 0.74 & -0.26 &$0.135\pm0.183$&$...$& 19.76 \\
\enddata
\tablenotetext{a}{{L}og effective column density defined as $\log \mnhi$ + [M/H].}
\end{deluxetable*}

\subsection{Constraining the Extinction and Reddening} \label{subsec:constraining_extinction}

The value of $\delta\alpha$ found in the previous section
indirectly reveals the average flux absorbed by the cMSDLAs sample
via dust reddening.  In this section we use $\delta\alpha$ to
estimate the average extinction ($A_\lambda$) and reddening ($E_{B-V}$).

Total extinction at a given wavelength is defined as:
\begin{equation}
A_\lambda = - 2.5 \cdot \log_{10} {f_{obs}(\lambda) \overwithdelims [] f_{intrinsic}(\lambda)}.
\end{equation}
We also wish to constrain the reddening 
$E_{B-V} \equiv A_B-A_V$, but we do not have wavelength 
coverage of the B and V bands in the absorber rest frame.  
Instead, we take a reddening value in the ultraviolet $E_{\lambda_1-\lambda_2} \equiv A_{\lambda_1}-A_{\lambda_2}$
and then use an assumed extinction law $\xi_\lambda$ to convert $E_{\lambda_1-\lambda_2}$
into $E_{B-V}$.

The control sample of QSOs with no absorption systems
give an intrinsic flux $f_{intrinsic}(\lambda)$ where no light is absorbed by dust.  The sample of cMSDLAs
represent systems where flux is lost to dust absorption $f_{obs}(\lambda)$.
We model the flux of the QSO continuum as the power-law seen in Equation \ref{eqtn:powerlaw}.
%$\begin{equation}
%4f(\lambda) = f_{obs}(1700\AA) \cdot {\lambda_{em} \overwithdelims () 1700 \AA}^{-\alpha}.
%$\end{equation}
%The model continuum for both the MSDLA sample and the control sample is
%based on the median $\alpha$ for each sample.
%The difference in $\alpha$ between the MSDLA sample and control sample is
%given by $\delta\alpha$.
For two wavelengths $\lambda_1$ and $\lambda_2$, the reddening
$E_{\lambda_1-\lambda_2}$ may be written as

\begin{eqnarray}
E_{\lambda_1-\lambda_2} &= A_{\lambda_1} - A_{\lambda_2} \\ 
\\
&= - 2.5 \cdot log_{10} \left [ {\lambda_1 \overwithdelims () \lambda_2}^{\delta\alpha} \right ] .
\nonumber 
\end{eqnarray}
To evaluate the reddening, we set $\lambda_1$ and $\lambda_2$ to values
representative of the emission-free windows in our analysis of the 
quasar continuum.
The analysis typically covers $\approx 1300 - 2250$\AA\ in the
quasars' rest-frame.  
Converting from the median QSO redshift of $z_{em}=2.087$ to the 
median absorber redshift of $z_{abs}=1.955$ gives 
$\lambda_1 \approx 1390$\AA\ and $\lambda_2 \approx 2400$\AA.  
For $\delta\alpha = 0.29$, we infer 
$E_{\lambda_1-\lambda_2} \approx 0.17$\, mag.

To convert to $E_{B-V}$, we must assume an extinction law.  Observations
to date suggest that DLAs exhibit SMC-like extinction \citep{ykv+06, wh+05}
and we adopt this as our default extinction law (and test this
assumption in the following sub-section).  With the SMC law, we 
infer $E_{B-V} \approx 0.025$ mag and an average $V$-band extinction
$A_V = E_{B-V} \cdot R_V \approx 0.076$ mag where we assumed $R_V = 3.1$.
The latter is the favored value for the Galaxy,
but also roughly holds for the SMC.
This $E_{B-V}$ value is systematically higher than any previously estimated
for the DLAs \citep{ml04,ehl05,vpw08} as expected for a sample of absorbers
pre-selected to have larger metal column densities.

\begin{deluxetable}{lll}
\tablewidth{0pc}
\tabletypesize{\footnotesize}
\tablecaption{Comparison of Dust-to-Gas Ratios \label{tab:dusttogasratio}}
\tablehead{
\colhead{Object(s)}&\colhead{$\log  ( A_V / N_{HI} )$}&\colhead{Reference}\\
}
\startdata
candidate MSDLAs & $\approx -21.97$ & \\
SMC  & $=-22.11$ & \cite{gcm+03}\\
Milky Way & $= -21.28$ & \cite{bohlin78} \\
248 DLAs & $\approx -22.53$ to $-22.40$ & \cite{vpw08}\\
\enddata
\end{deluxetable}

With our predicted values for extinction and reddening, we can estimate 
the dust-to-gas ratio for the cMSDLAs.  The cMSDLAs have a median \ion{H}{1} 
column density of  log \nhi\ = 20.85 dex giving a median dust-to-gas ratio of 
$\log  ( A_V / \mnhi ) \approx -22.0$ for the SMC extinction law.  
We ignore the contribution of molecular hydrogen because it is expected to be a negligible contribution to the total amount of hydrogen ($\S$~\ref{sect:metal}).
A comparison to dust-to-gas ratios 
in the literature can be seen in Table \ref{tab:dusttogasratio}.
Our value for the dust-to-gas ratio of cMSDLAs is very similar 
to the SMC value
$\log  ( A_V / \mnhi ) = -22.1$ \citep{gcm+03}, with 
a difference of only 0.1\,dex.  
%It should be noted that the predicted value for $\log  ( A_V / N_{HI} )$ 
%is based on an assumed SMC extinction law,
%%Added following sentences in as suggested by Stephane, but need some double checking i think.
%although we are uncertain if our assumed extinction law(s) are biasing the result.
%This is important to note, especially since our 
%predicted dust-to-gas ratio is so close to the value for the SMC.
\cite{vcl+06} suggests that the dust-to-gas ratio scales with metallicity.  
We have metallicity values 
for 22 of the cMSDLAs, giving a median metallicity of [M/H]~$\approx -0.6$.  
This roughly follows the expected trend of dust-to-gas ratio vs.\ 
metallicity suggested by \cite{vcl+06}.
We conclude, therefore, that the dust-to-gas properties
of the MSDLAs are generally consistent
with that observed in the SMC.  We also note that these
inferences are subject to the assumed extinction law.

\subsection{The Possibility of the 2175 \AA\ Bump}

To test the assumption of an SMC-like extinction law, we have searched 
for the 2175 \AA\ bump characteristic of a Galactic
extinction law.
The exact nature of the 2175 \AA\ bump is unclear.  It is thought to arise from the fine structure transitions of Polycyclic Aromatic Hydrocarbon (PHC) molecules in the Milky Way Galaxy \citep{a+h03}.
The detection of the  2175 \AA\ bump in the cMSDLAs would infer the existance of PHCs at high redshift \citep{wang04}.
A visual inspection for a broad absorption feature at 2175 \AA\ in the MMT spectra for each of the cMSDLAs yields no identifiable features.  
\cite{malhotra97} was the first to attempt this sort of analysis.  She created 
a composite spectrum of 96 Mg II absorbers and claimed a
detection of the 2175 \AA\ bump.  
Subsequent studies, however, have not confirmed this result \citep{mzn+05}.
\cite{vpw08} and \cite{vcl+06} carried out similar searches using DLA samples.  
\cite{vpw08} found no conclusive evidence for a 2175\AA\ bump in their data while 
\cite{vcl+06} found evidence for a 2175\AA\ bump in 1 out of 8 possible systems
pre-selected to have very strong metal absorption.  
Although rare in DLAs, it is important to search for the 2175\AA\ bump to test 
our use of an SMC-like extinction law.
A composite spectrum of the cMSDLAs (shifted into the absorber rest frame) with a significantly higher signal-to noise ratio might yield evidence for a 2175 \AA\ bump.  

\begin{figure}
\begin{center}
\includegraphics[height=3.5in,angle=90]{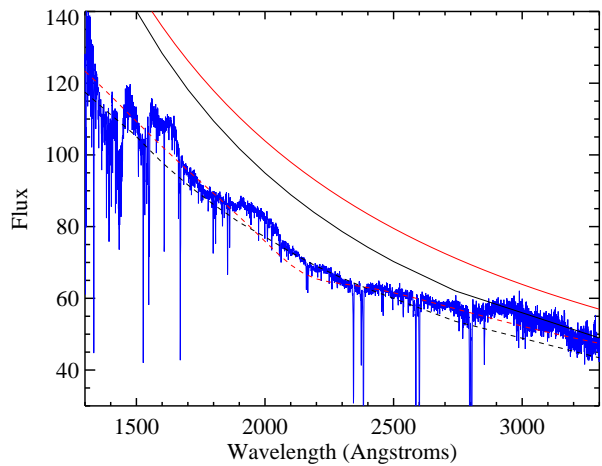}
\end{center}
\caption{
The composite spectrum of the cMSDLAs (blue).  Two extinction-law models are overplotted,  the SMC-like extinction law model (black) and the MW-like extinction law model (red) which contains the 2175\AA\ bump.  The solid lines above the composite spectrum are the power-law continua used for each model.  The dotted lines overlapping the composite spectrum are the results of artificially reddening the continua.  A possible broad absorption line feature is visible at $\sim$ 2135 \AA, but we are unable to confirm if this feature is genuine or caused by the variation in the composite spectrum's continuum. 
Therefore we claim no positive detection of the 2175\AA\ bump.
}
\label{fig:cmsdlacomposite}
\end{figure}

We constructed a composite spectrum of all our cMSDLAs by shifting each of their SDSS spectra into the DLA rest-frame, binning the pixels with 0.5 \AA\ bins, and finding the weighted mean for all the pixels in each bin.  We attempt to compensate for variation in the brightness of each QSO by having each individual spectrum scaled so that the median flux between 1000 and 1800 \AA\ (QSO rest-frame) equals 100.  The composite SDSS spectrum of the cMSDLAs can be seen in Figure \ref{fig:cmsdlacomposite}.  Examining this composite spectrum does yield a possible broad absorption feature at $\sim$ 2135 \AA. This identification is questionable because of the high amount of irregularity in the continuum of the composite spectrum.  This irregularity is caused by overlapping QSO broad emission lines from the individual spectra getting smeared out in the composite.  The relatively low number of systems stacked in the composite spectrum is not enough to completely smooth out the emission lines intrinsic to the QSOs, where normally a composite of hundreds of systems would be needed.  Two extinction law models are created, one for a MW-like extinction law and another for a SMC-like extinction law, which are overplotted with the composite cMSDLA spectrum.  For each extinction model, a continuum is modeled with a power-law ($f \varpropto \lambda^{-\alpha}$) which is then artificially reddened with the respective extinction law to match the composite spectrum.  These models can be seen in Figure \ref{fig:cmsdlacomposite}.  The results of these fits are inconclusive.  
We conclude that the assumption of an SMC-like extinction law
is justified although not required by our data.

%THIS INFORMATION IS HEARBY COSIGNED TO OBLIVION FOR NOW
%Figure \ref{fig:mwbump} shows a composite candidate MSDLA spectrum vs. a composite spectrum of many ordinary DLAs.
%There is a possible broad absorption feature observed in the composite spectrum of MSDLA candidates with a centroid at $\sim$ 2135 \AA\ and a FWHM of $\sim$ 130 \AA.  The depth of the possible absorption feature is $\sim$ 0.04 normalized flux.  This is in contrast to \cite{malhotra97} who gives a centroid at $\sim$ 2240 \AA\ and a FWHM of 200-300 \AA.  Attempts to fit artificially reddened model continua to both the candidate MSDLA and ordinary DLA composite spectra were inconclusive, although detectable Milky Way like extinction is ruled out for the ordinary DLA composite.  The evidence for a 2175 \AA\ bump is inconclusive for these absorption systems.  The most likely explanation for why there appears to be a broad absorption feature in the candidate MSDLA composite spectrum is low numbers where there were not enough objects in the sample making up the composite to smooth out large features such as QSO Broad Emission Lines (BELs).  A future composite of MSDLA spectra with QSO BELs masked out could help answer these questions.

\subsection{Extinction Per Unit Column Density of Iron in Candidate MSDLAs}

Previous works have asserted that gas in the DLAs and the ISM of local
galaxies conforms to a nearly constant extinction per column density
of metals in the dust phase \citep{vcl+06}.  
We can further explore this assertion with out sample of DLA systems.
Specifically, the 
% N_Zn = [M/H] +N_HI - 12 + 4.67
extinction per unit column density of iron in the 
dust phase, $\left<s_V^{Fe}\right>$, is given by:
\begin{equation}
\left< s_V^{Fe} \right> = {A_V \over f_{Fe} \widehat{N}_{Fe} } \quad , 
%= %{A_V \over f_{Fe} 
%\left( \left [ {M \over H} \right] + N_{HI} - 7.33 \right)
%N_{Zn}
% \left( {Fe \over Zn} \right)_\odot},
\end{equation}
where $\widehat{N}_{Fe}$ is the dust phase column density of iron and $f_{Fe}$ is the fraction of total iron atoms in the dust.
The value of $\widehat{N}_{Fe}$ is often estimated by assuming Fe/Zn
has an intrinsic, solar relative abundance, i.e.
$\widehat{N}_{\rm Fe} = N_{\rm Zn} \rm (Fe/Zn)_\odot$,
assuming Zn has a negligible depletion into dust grains.
If the extinction properties of dust are the same throughout the universe, one would expect that $\widehat{N}_{Fe} \propto A_V$ across a range of redshifts and metallicities.  If this is true, then  $\left< s_V^{Fe} \right>$ is expected to be similar for any sight line, whether in our the Galaxy or in high redshift DLAs \citep{vcl+06}.

We proceed assuming a solar 
iron-to-zinc abundance ratio $\left( {Fe \over Zn } \right)_\odot = 2.94$ 
\citep{asplund09} for the cMSDLAs and that all of the iron 
is in the dust phase ($f_{Fe} = 1$).
Of the 43 cMSDLAs, twenty-two have previous metallicity measurements with
an average zinc column density of 
$\log N_{Zn} = \left [ {M \over H} \right ] + \log N_{\rm HI} - 7.33= 12.95$ dex. 
With $A_V \approx 0.076$\,mag, 
we obtain $\left < s_V^{Fe} \right > \approx 1 \times 10^{-17}$ mag cm$^2$ for the cMSDLA sample.   
This result is comparable to the value estimated by \cite{vcl+06}
for a set of similar systems.
It further supports their assertion that dust in the DLAs exhibits
a roughly average extinction per column density of metals.
In turn, this result stresses that the MSDLAs are excellent candidates
to examining reddening by the ISM of high $z$ galaxies.

\subsection{Dust Absorption for Individual Systems}

While we have measured the average extinction and reddening of the
cMSDLAs, we would like to measure absorption by dust in each 
individual system.  This is challenging 
because the intrinsic QSO continua are unknown
and likely exhibit large scatter.  
The previous analysis shows that $\alpha$ is depressed for the cMSDLAs 
and the power-law used to model the QSO continuum does provide a lower 
limit on the intrinsic flux which can be used.  We use the previous power-law fits to extrapolate the QSO continuum further into the blue, specifically in the region between the QSO's \lya\ and \lyb\ emission lines.  
Departures from a strict power-law may indicate dust extinction and,
in principle, could yield the extinction law for individual galaxies
at far-UV wavelengths.
In practice, however, our uncertainty on the intrinsic spectrum and
(more importantly) absorption by the \lya\ forest 
complicates the analysis.  Nevertheless, this presents a test of our
previous analysis from the power-law slopes measured outside the \lya\ 
forest ($\S$~\ref{subsec:powerlawslope}).

%Information on Delta-Alpha
The continuum of a QSO redward of $\approx 1300$\AA\ is well characterized 
by a single power-law $f_r(\lambda) \propto \lambda^{\alpha_r}$.  
In the far-UV, however, there is believed to be a break in 
this power-law to a second power-law $f_b(\lambda) \propto \lambda^{\alpha_b}$.  
We define the difference in exponents between these 
two power-laws as $\mdeltaalpha \equiv \alpha_r - \alpha_b$.   
Composite QSO HST spectra by \cite{telfer02} gives 
\deltaalpha$=1.07\pm0.14$ with a break at $\approx$1250 \AA.  
A similar analysis by \cite{zheng1997} gives \deltaalpha=0.97$\pm$0.16 
with a break at 1050 \AA.
%Unsure if I will include or not include the following sentences:
%The uncertainty in where break occurs is at least partially caused by the QSO's \lya\ emission line obscuring the continuum in this region.  Whether the \lya\ line is part of the physical mechanism causing the break is unknown.
For our analysis, we adopt a value of \deltaalpha=1.0 with a break at the QSO's \lya\ emission line (1216 \AA).

For each quasar, we
create a model continuum by fitting to emission-free 
regions in the SDSS spectrum ($\S$~\ref{subsec:powerlawslope}).  
The shape of the continuum redward of 
1216\AA\ is described by a power-law with the exponent 
$\alpha_r$.  This initial power-law breaks blueward of 
1216\AA\ and becomes a second power-law with 
$\alpha_b=\alpha_r-\Delta\alpha$, such as: 
\begin{equation}
f_b(\lambda<1216{\rm \AA})=f(1216 {\rm \AA}){\lambda \overwithdelims () 
    1216 {\rm \AA}}^{-\alpha_b}
\end{equation}

Our model QSO continuum is derived from the SDSS spectrum because it is assumed to have accurate flux and covers redder wavelengths where extinction is 
reduced.  By extrapolating the model continuum into the blue, we aim 
to place a lower limit on the dust absorption in each system.  
We use the wavelengths between the QSO's \lya\ and \lyb\ emission lines 
to measure the dust absorption (1070 to 1170 \AA).  
In this way the results are not contaminated by the QSO's emission lines.  
%The difficulty with this spectral range is 
%absorption from intergalactic neutral hydrogen, i.e.\ the \lya\ forest.  
To estimate the contribution of the \lya\ forest to the absorbed flux,
we used values for the average absorption by the \lya\ forest (DA) provided 
by \cite{kts+05}:
\begin{equation}
DA = 0.0062 (1+z)^{2.75}.
\end{equation}
The flux in the region of extrapolated QSO continuum between the QSO's 
\lya\ and \lyb\ emission lines is then multiplied by DA to scale it down 
to the estimated flux after \lya\ forest 
absorption.  To avoid contamination from the DLA's \lya\ absorption line, each line is individually masked out.   
Furthermore, the MMT spectrum blueward of 
3190\AA\ is avoided because of excessive noise.  
The mean flux of the \lya\ to \lyb\ region in 
both the SDSS and MMT spectra is then compared with the mean flux of the same region in the model continuum.  If the mean flux of the model continuum is higher than the mean flux of the actual continuum, we attribute the lower than expected flux to absorption by dust in the DLA.

The fraction of observed flux compared to flux predicted by the model QSO continuum is termed \textit{d} and is calculated as follows,
\begin{equation}
\textit{d} = 1-{\overline{f}_{obs}(\lambda) \over 
\overline{f}_b(\lambda)\cdot(1-DA)} \;\;\; ,
\end{equation}  
where $\overline{f}_{obs}$ is the average observed flux between 
1070 and 1170\AA\, and is defined as:
\begin{equation}
\overline{f}_{obs}(\lambda) = {{\sum_{i=1070\rAA}^{1170 \rAA} f_{obs}^i(\lambda_{obs}) \cdot \Delta\lambda_{obs}^i} \over 
                                   {\sum_{i=1070\rAA}^{1170 \rAA} \Delta\lambda_{obs}^i}} \;\;\; .
\end{equation}
Finally,
$\overline{f}_b(\lambda)$ is the average flux of the model 
QSO continuum between 1070 to 1170 \AA\ and is given by
\begin{equation}
\overline{f}_b(\lambda) =  {1 \over \Delta\lambda}  \int\displaylimits_{1070 \rAA}^{1170 \rAA} f_b(\lambda) d\lambda .
\end{equation}
(1-DA) is the predicted fractional throughput of flux after absorption by the \lya\ forest.
The individual results for \textit{d} can be seen in Table \ref{tab:dust}.  If the parameter \textit{d} truly scales with the amount of dust in a system, one would expect it is correlated with the system's measured metal column density.  

\begin{figure}
\begin{center}
\includegraphics[height=3.5in,angle=90]{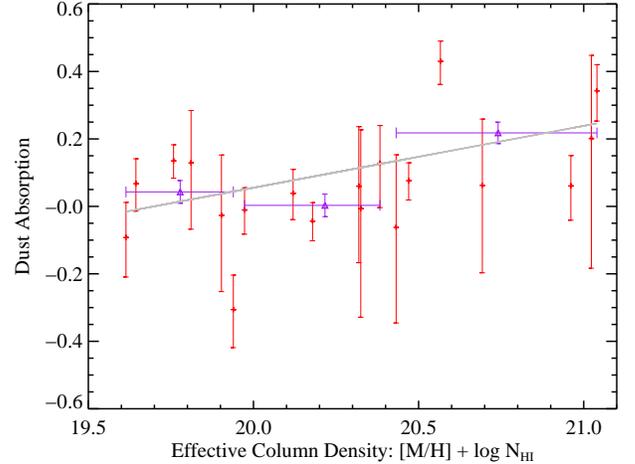}
\end{center}
\caption{
Dust absorption (as defined in the text) versus
the effective column density 
of each cMSDLA with a measured metal column density and dust absorption value.  Each of the three bins show the weighted mean of six DLAs.  The regression line fit to the data has a slope of 0.185 $\pm$ 0.046.  A bootstrap analysis of 10,000 random samplings of the dataset indicates
a positive slope at 95\% confidence limit,
suggesting reddening from dust 
within the cMSDLAs in the far-UV.
The purple points are bins that each represent the weighted mean of $\sim 1/3$ of the dataset.
}
\label{fig:dustabs}
\end{figure}

For the latter, we introduce an effective column density 
$\log N_{\rm eff} \equiv \log \mnhi + {\rm [M/H]}$ that
characterizes the total metal column density for the absorber.
We plot \textit{d} vs.\ \neff\ in Figure \ref{fig:dustabs}.  
A regression line fit to the data gives a slope of 0.185 $\pm$ 0.046 
(statistical error). 
A bootstrap analysis of 10,000 random samplings of the dataset indicates
a positive slope at 95\% confidence limit. 
The positive correlation between \textit{d} and \neff\
serves as evidence that MSDLAs contain 
detectable, if not significant, amounts of dust.

Future work will focus on these systems.  By obtaining spectrophotometry
from optical to near-IR wavelengths, we can constrain the intrinsic slope
of the quasar and better estimate the underlying extinction law.

\section{Summary and Concluding Remarks}
\label{sec:summary}

We obtained MMT/BCS spectra of 41 QSO sightlines where previous observations found high metal column densities.  We classify these absorption systems as candidate (c)MSDLAs.  
49 absorbers with strong \lya\ lines ($W > 5$\AA)
were identified, 43 of which correspond to the targeted cMSDLAs.  
The redshift for all absorption systems was determined visually by finding the centroid of low ion metal transitions.  The H I column density (\ncol) for all absorption systems was measured by visually fitting a Voigt profile to the absorber's \lya\ line.  Out of 49 absorbers, 42 met the DLA criterion (\ncol$\ge 20.3$ \atomspercm)
including all but two of the targeted cMSDLAs.  
The MSDLAs and cMSDLA show higher systematic H I column densities than comparable systems, in line with the expectation that selecting systems with high metal column density will also select systems with high H I column density.   

Using the measured H I column densities and previously measured metal column densities, the metallicity [M/H] for each system was calculated.  MSDLAs and cMSDLAs exhibit systematically higher metallicities than DLAs that are not metal strong.  This suggests that MSDLA sightlines might probe towards the centers of galaxies where metallicity is greater, assuming a metallicity gradient.  Perhaps MSDLAs represent a population of high redshift galaxies that became chemically enriched earlier in their evolution than others.  Using these new calculations, we examined and extended
the metallicity vs.\ \ion{Si}{2} 1526 EW relation 
discussed by \cite{pcw+08}.  If this correlation is set by a underlying
mass/metallicity relation, then the MSDLAs may represent the highest
mass, gas-rich galaxies at high $z$.

We searched for signatures of dust in the cMSDLAs by studying the reddening of the
background quasars.  A power-law fit to the continuum of the SDSS spectra yielded individual power-laws (parameterized by $\alpha$)
that model the continuum of each QSO.  
The quasars behind cMSDLAs have lower $\alpha$'s then a
control sample ($\delta\alpha = 0.29$) indicating significant
extinction by dust in the cMSDLAs.
Assuming an SMC-like extinction law, we estimate 
$A_V \approx 0.076$\,mag and $E_{B-V} \approx 0.025$ for these
systems.  
The dust-to-gas ratio for the cMSDLAs was estimated 
to be $\log  ( A_V / N_{HI} ) \approx -22.0$, 
similar to the ratio for the SMC and roughly following the 
expected trend of the dust-to-gas ratio scaling with metallicity.   
The ratio of extinction to dust phase iron column density was estimated to be  $\left<s_V^{Fe} \right> \approx 1 \times 10^{-17}$.  

%We attempted to measure the fraction of flux from the background QSOs absorbed by dust for each individual DLA.  We created a model QSO spectrum based on the initial power-law fits and compared the mean flux between the QSO's \lya\ and \lyb\ emission lines to our predicted intrinsic flux in that region.  If we compare our values for flux lost due to dust absorption to metal column density for each system, one would expect a correlation between the two if metal column density traces the amount of dust in a system.  We do find a positive relation suggesting that MSDLAs contain dust which absorb the light of the background QSOs.  MSDLAs do not seem to just be DLAs which contain high metal column densities.  
The MSDLAs represent a unique subset of DLAs which are more chemically enriched, massive, dusty, and possibly evolved then ordinary DLAs.
As such, they are excellent targets for future studies of
extinction, nucleosynthesis, and searches for stellar light and 
molecular emission lines at high $z$.

%Not quite sure where this should go but most other papers put them as the last paragraph so I did the same
\acknowledgements
KFK was supported by an NSF REU grant related to AST-0709235.
JXP is supported by NSF grant (AST-0709235).
Observations reported here were obtained at the MMT Observatory, 
a joint facility of the University of Arizona and the Smithsonian Institution.

%\bibliographystyle{/u/xavier/paper/Bibli/apj}
%\bibliography{/u/xavier/paper/Bibli/allrefs}

%\bibliographystyle{/Users/Kyle/Bibli/apj}
%\bibliography{/Users/Kyle/Bibli/allrefs}

\begin{thebibliography}{48}
\expandafter\ifx\csname natexlab\endcsname\relax\def\natexlab#1{#1}\fi



\bibitem[{{Adelman-McCarthy} {et~al.}(2008){Adelman-McCarthy}, {Ag{\"u}eros},
  {Allam}, {Allende Prieto}, {Anderson}, {Anderson}, {Annis}, {Bahcall},
  {Bailer-Jones}, {Baldry}, {Barentine}, {Bassett}, {Becker}, {Beers}, {Bell},
  {Berlind}, {Bernardi}, {Blanton}, {Bochanski}, {Boroski}, {Brinchmann},
  {Brinkmann}, {Brunner}, {Budav{\'a}ri}, {Carliles}, {Carr}, {Castander},
  {Cinabro}, {Cool}, {Covey}, {Csabai}, {Cunha}, {Davenport}, {Dilday}, {Doi},
  {Eisenstein}, {Evans}, {Fan}, {Finkbeiner}, {Friedman}, {Frieman},
  {Fukugita}, {G{\"a}nsicke}, {Gates}, {Gillespie}, {Glazebrook}, {Gray},
  {Grebel}, {Gunn}, {Gurbani}, {Hall}, {Harding}, {Harvanek}, {Hawley},
  {Hayes}, {Heckman}, {Hendry}, {Hindsley}, {Hirata}, {Hogan}, {Hogg}, {Hyde},
  {Ichikawa}, {Ivezi{\'c}}, {Jester}, {Johnson}, {Jorgensen}, {Juri{\'c}},
  {Kent}, {Kessler}, {Kleinman}, {Knapp}, {Kron}, {Krzesinski}, {Kuropatkin},
  {Lamb}, {Lampeitl}, {Lebedeva}, {Lee}, {Leger}, {L{\'e}pine}, {Lima}, {Lin},
  {Long}, {Loomis}, {Loveday}, {Lupton}, {Malanushenko}, {Malanushenko},
  {Mandelbaum}, {Margon}, {Marriner}, {Mart{\'{\i}}nez-Delgado}, {Matsubara},
  {McGehee}, {McKay}, {Meiksin}, {Morrison}, {Munn}, {Nakajima}, {Neilsen},
  {Newberg}, {Nichol}, {Nicinski}, {Nieto-Santisteban}, {Nitta}, {Okamura},
  {Owen}, {Oyaizu}, {Padmanabhan}, {Pan}, {Park}, {Peoples}, {Pier}, {Pope},
  {Purger}, {Raddick}, {Re Fiorentin}, {Richards}, {Richmond}, {Riess}, {Rix},
  {Rockosi}, {Sako}, {Schlegel}, {Schneider}, {Schreiber}, {Schwope}, {Seljak},
  {Sesar}, {Sheldon}, {Shimasaku}, {Sivarani}, {Smith}, {Snedden}, {Steinmetz},
  {Strauss}, {SubbaRao}, {Suto}, {Szalay}, {Szapudi}, {Szkody}, {Tegmark},
  {Thakar}, {Tremonti}, {Tucker}, {Uomoto}, {Vanden Berk}, {Vandenberg},
  {Vidrih}, {Vogeley}, {Voges}, {Vogt}, {Wadadekar}, {Weinberg}, {West},
  {White}, {Wilhite}, {Yanny}, {Yocum}, {York}, {Zehavi}, \&
  {Zucker}}]{sdssdr6}
{Adelman-McCarthy}, J.~K., {et~al.} 2008, \apjs, 175, 297

\bibitem[{{Allamandola} \& {Hudgins}(2003)}]{a+h03}
{Allamandola}, L.~J., \& {Hudgins}, D.~M. 2003, in Solid State Astrochemistry,
  ed. {V.~Pirronello, J.~Krelowski, \& G.~Manic{\`o}}, 251--316

\bibitem[{{Asplund} {et~al.}(2009){Asplund}, {Grevesse}, {Sauval}, \&
  {Scott}}]{asplund09}
{Asplund}, M., {Grevesse}, N., {Sauval}, A.~J., \& {Scott}, P. 2009, ArXiv
  e-prints

\bibitem[{{Bohlin} {et~al.}(1978){Bohlin}, {Savage}, \& {Drake}}]{bohlin78}
{Bohlin}, R.~C., {Savage}, B.~D., \& {Drake}, J.~F. 1978, \apj, 224, 132

\bibitem[{{Boisse} {et~al.}(1998){Boisse}, {Le Brun}, {Bergeron}, \&
  {Deharveng}}]{blb+98}
{Boisse}, P., {Le Brun}, V., {Bergeron}, J., \& {Deharveng}, J.-M. 1998, \aap,
  333, 841

\bibitem[{{Ellison} {et~al.}(2005){Ellison}, {Hall}, \& {Lira}}]{ehl05}
{Ellison}, S.~L., {Hall}, P.~B., \& {Lira}, P. 2005, \aj, 130, 1345

\bibitem[{{Erb} {et~al.}(2006){Erb}, {Shapley}, {Pettini}, {Steidel}, {Reddy},
  \& {Adelberger}}]{esp+06}
{Erb}, D.~K., {Shapley}, A.~E., {Pettini}, M., {Steidel}, C.~C., {Reddy},
  N.~A., \& {Adelberger}, K.~L. 2006, \apj, 644, 813

\bibitem[{{Ge} \& {Bechtold}(1997)}]{gb97}
{Ge}, J., \& {Bechtold}, J. 1997, \apjl, 477, L73+

\bibitem[{{Gordon} {et~al.}(2003){Gordon}, {Clayton}, {Misselt}, {Landolt}, \&
  {Wolff}}]{gcm+03}
{Gordon}, K.~D., {Clayton}, G.~C., {Misselt}, K.~A., {Landolt}, A.~U., \&
  {Wolff}, M.~J. 2003, \apj, 594, 279

\bibitem[{{Hamann} {et~al.}(2008){Hamann}, {Kaplan}, {Rodriguez Hidalgo},
  {Prochaska}, \& {Herbert-Fort}}]{hkr+08}
{Hamann}, F., {Kaplan}, K.~F., {Rodriguez Hidalgo}, P., {Prochaska}, J.~X., \&
  {Herbert-Fort}, S. 2008, ArXiv e-prints

\bibitem[{{Herbert-Fort} {et~al.}(2006){Herbert-Fort}, {Prochaska},
  {Dessauges-Zavadsky}, {Ellison}, {Howk}, {Wolfe}, \& {Prochter}}]{shf06}
{Herbert-Fort}, S., {Prochaska}, J.~X., {Dessauges-Zavadsky}, M., {Ellison},
  S.~L., {Howk}, J.~C., {Wolfe}, A.~M., \& {Prochter}, G.~E. 2006, \pasp, 118,
  1077

\bibitem[{{Jenkins}(2009)}]{jenkins09}
{Jenkins}, E.~B. 2009, \apj, 700, 1299

\bibitem[{{Kirkman} {et~al.}(2005){Kirkman}, {Tytler}, {Suzuki}, {Melis},
  {Hollywood}, {James}, {So}, {Lubin}, {Jena}, {Norman}, \& {Paschos}}]{kts+05}
{Kirkman}, D., {et~al.} 2005, \mnras, 360, 1373

\bibitem[{{Krumholz} {et~al.}(2009){Krumholz}, {Ellison}, {Prochaska}, \&
  {Tumlinson}}]{kep+09}
{Krumholz}, M.~R., {Ellison}, S.~L., {Prochaska}, J.~X., \& {Tumlinson}, J.
  2009, \apjl, 701, L12

\bibitem[{{Lu} {et~al.}(1996){Lu}, {Sargent}, {Barlow}, {Churchill}, \&
  {Vogt}}]{lu96}
{Lu}, L., {Sargent}, W.~L.~W., {Barlow}, T.~A., {Churchill}, C.~W., \& {Vogt},
  S.~S. 1996, \apjs, 107, 475

\bibitem[{{Malhotra}(1997)}]{malhotra97}
{Malhotra}, S. 1997, \apjl, 488, L101+

\bibitem[{{M{\'e}nard} {et~al.}(2005){M{\'e}nard}, {Zibetti}, {Nestor}, \&
  {Turnshek}}]{mzn+05}
{M{\'e}nard}, B., {Zibetti}, S., {Nestor}, D., \& {Turnshek}, D. 2005, in IAU
  Colloq. 199: Probing Galaxies through Quasar Absorption Lines, ed.
  P.~{Williams}, C.-G. {Shu}, \& B.~{Menard}, 86--91

\bibitem[{{M{\"o}ller} {et~al.}(2004){M{\"o}ller}, {Fynbo}, \& {Fall}}]{mff04}
{M{\"o}ller}, P., {Fynbo}, J.~P.~U., \& {Fall}, S.~M. 2004, \aap, 422, L33

\bibitem[{{Murphy} {et~al.}(2007){Murphy}, {Curran}, {Webb}, {Menager}, \&
  {Zych}}]{mcw+07}
{Murphy}, M.~T., {Curran}, S.~J., {Webb}, J.~K., {Menager}, H., \& {Zych},
  B.~J. 2007, ArXiv Astrophysics e-prints

\bibitem[{{Murphy} \& {Liske}(2004)}]{ml04}
{Murphy}, M.~T., \& {Liske}, J. 2004, \mnras, 354, L31

\bibitem[{{Nagamine} {et~al.}(2004){Nagamine}, {Springel}, \&
  {Hernquist}}]{nsh04}
{Nagamine}, K., {Springel}, V., \& {Hernquist}, L. 2004, \mnras, 348, 421

\bibitem[{{Noterdaeme} {et~al.}(2008){Noterdaeme}, {Ledoux}, {Petitjean}, \&
  {Srianand}}]{nlp+08}
{Noterdaeme}, P., {Ledoux}, C., {Petitjean}, P., \& {Srianand}, R. 2008, \aap,
  481, 327

\bibitem[{{O'Meara} {et~al.}(2007){O'Meara}, {Prochaska}, {Burles}, {Prochter},
  {Bernstein}, \& {Burgess}}]{opb+07}
{O'Meara}, J.~M., {Prochaska}, J.~X., {Burles}, S., {Prochter}, G.,
  {Bernstein}, R.~A., \& {Burgess}, K.~M. 2007, \apj, 656, 666

\bibitem[{{Ostriker} \& {Heisler}(1984)}]{oh84}
{Ostriker}, J.~P., \& {Heisler}, J. 1984, \apj, 278, 1

\bibitem[{{Pei} {et~al.}(1991){Pei}, {Fall}, \& {Bechtold}}]{pfb91}
{Pei}, Y.~C., {Fall}, S.~M., \& {Bechtold}, J. 1991, \apj, 378, 6

\bibitem[{{P{\'e}roux} {et~al.}(2006){P{\'e}roux}, {Kulkarni}, {Meiring},
  {Ferlet}, {Khare}, {Lauroesch}, {Vladilo}, \& {York}}]{peroux06}
{P{\'e}roux}, C., {Kulkarni}, V.~P., {Meiring}, J., {Ferlet}, R., {Khare}, P.,
  {Lauroesch}, J.~T., {Vladilo}, G., \& {York}, D.~G. 2006, \aap, 450, 53

\bibitem[{{Petitjean} {et~al.}(2000){Petitjean}, {Srianand}, \&
  {Ledoux}}]{petit00}
{Petitjean}, P., {Srianand}, R., \& {Ledoux}, C. 2000, \aap, 364, L26

\bibitem[{{Pettini} {et~al.}(1994){Pettini}, {Smith}, {Hunstead}, \&
  {King}}]{pshk94}
{Pettini}, M., {Smith}, L.~J., {Hunstead}, R.~W., \& {King}, D.~L. 1994, \apj,
  426, 79

\bibitem[{{Pontzen} {et~al.}(2008){Pontzen}, {Governato}, {Pettini}, {Booth},
  {Stinson}, {Wadsley}, {Brooks}, {Quinn}, \& {Haehnelt}}]{pgp+08}
{Pontzen}, A., {et~al.} 2008, \mnras, 390, 1349

\bibitem[{{Prochaska} {et~al.}(2008){Prochaska}, {Chen}, {Wolfe},
  {Dessauges-Zavadsky}, \& {Bloom}}]{pcw+08}
{Prochaska}, J.~X., {Chen}, H.-W., {Wolfe}, A.~M., {Dessauges-Zavadsky}, M., \&
  {Bloom}, J.~S. 2008, \apj, 672, 59

\bibitem[{{Prochaska} {et~al.}(2003{\natexlab{a}}){Prochaska}, {Gawiser},
  {Wolfe}, {Castro}, \& {Djorgovski}}]{pgw+03}
{Prochaska}, J.~X., {Gawiser}, E., {Wolfe}, A.~M., {Castro}, S., \&
  {Djorgovski}, S.~G. 2003{\natexlab{a}}, \apjl, 595, L9

\bibitem[{{Prochaska} {et~al.}(2003{\natexlab{b}}){Prochaska}, {Gawiser},
  {Wolfe}, {Cooke}, \& {Gelino}}]{p03_esi}
{Prochaska}, J.~X., {Gawiser}, E., {Wolfe}, A.~M., {Cooke}, J., \& {Gelino}, D.
  2003{\natexlab{b}}, \apjs, 147, 227

\bibitem[{{Prochaska} {et~al.}(2005){Prochaska}, {Herbert-Fort}, \&
  {Wolfe}}]{phw05}
{Prochaska}, J.~X., {Herbert-Fort}, S., \& {Wolfe}, A.~M. 2005, \apj, 635, 123

\bibitem[{{Prochaska} {et~al.}(2003{\natexlab{c}}){Prochaska}, {Howk}, \&
  {Wolfe}}]{phw03}
{Prochaska}, J.~X., {Howk}, J.~C., \& {Wolfe}, A.~M. 2003{\natexlab{c}}, \nat,
  423, 57

\bibitem[{{Prochaska} {et~al.}(2006){Prochaska}, {O'Meara}, {Herbert-Fort},
  {Burles}, {Prochter}, \& {Bernstein}}]{pho+06}
{Prochaska}, J.~X., {O'Meara}, J.~M., {Herbert-Fort}, S., {Burles}, S.,
  {Prochter}, G.~E., \& {Bernstein}, R.~A. 2006, \apjl, 648, L97

\bibitem[{{Prochaska} \& {Wolfe}(1996)}]{pw96}
{Prochaska}, J.~X., \& {Wolfe}, A.~M. 1996, \apj, 470, 403

\bibitem[{{Prochaska} \& {Wolfe}(2001)}]{pw01}
---. 2001, \apjl, 560, L33

\bibitem[{{Prochaska} \& {Wolfe}(2009)}]{pw09}
---. 2009, \apj, 696, 1543

\bibitem[{{Prochaska} {et~al.}(2007){Prochaska}, {Wolfe}, {Howk}, {Gawiser},
  {Burles}, \& {Cooke}}]{pwh+07}
{Prochaska}, J.~X., {Wolfe}, A.~M., {Howk}, J.~C., {Gawiser}, E., {Burles},
  S.~M., \& {Cooke}, J. 2007, \apjs, 171, 29

\bibitem[{{Prochaska} {et~al.}(2001){Prochaska}, {Wolfe}, {Tytler}, {Burles},
  {Cooke}, {Gawiser}, {Kirkman}, {O'Meara}, \& {Storrie-Lombardi}}]{pro01}
{Prochaska}, J.~X., {et~al.} 2001, \apjs, 137, 21

\bibitem[{{Savage} \& {Mathis}(1979)}]{savage+mathis79}
{Savage}, B.~D., \& {Mathis}, J.~S. 1979, \araa, 17, 73

\bibitem[{{Savage} \& {Sembach}(1996)}]{ss96}
{Savage}, B.~D., \& {Sembach}, K.~R. 1996, \araa, 34, 279

\bibitem[{{Telfer} {et~al.}(2002){Telfer}, {Zheng}, {Kriss}, \&
  {Davidsen}}]{telfer02}
{Telfer}, R.~C., {Zheng}, W., {Kriss}, G.~A., \& {Davidsen}, A.~F. 2002, \apj,
  565, 773

\bibitem[{{Vladilo} {et~al.}(2006){Vladilo}, {Centuri{\'o}n}, {Levshakov},
  {P{\'e}roux}, {Khare}, {Kulkarni}, \& {York}}]{vcl+06}
{Vladilo}, G., {Centuri{\'o}n}, M., {Levshakov}, S.~A., {P{\'e}roux}, C.,
  {Khare}, P., {Kulkarni}, V.~P., \& {York}, D.~G. 2006, \aap, 454, 151

\bibitem[{{Vladilo} {et~al.}(2008){Vladilo}, {Prochaska}, \& {Wolfe}}]{vpw08}
{Vladilo}, G., {Prochaska}, J.~X., \& {Wolfe}, A.~M. 2008, \aap, 478, 701

\bibitem[{{Wang} {et~al.}(2004){Wang}, {Hall}, {Ge}, {Li}, \&
  {Schneider}}]{wang04}
{Wang}, J., {Hall}, P.~B., {Ge}, J., {Li}, A., \& {Schneider}, D.~P. 2004,
  \apj, 609, 589

\bibitem[{{Wild} \& {Hewett}(2005)}]{wh+05}
{Wild}, V., \& {Hewett}, P.~C. 2005, \mnras, 361, L30

\bibitem[{{Wolfe} {et~al.}(2005){Wolfe}, {Gawiser}, \& {Prochaska}}]{wgp05}
{Wolfe}, A.~M., {Gawiser}, E., \& {Prochaska}, J.~X. 2005, \araa, 43, 861

\bibitem[{{York} {et~al.}(2006){York}, {Khare}, {Vanden Berk}, {Kulkarni},
  {Crotts}, {Lauroesch}, {Richards}, {Schneider}, {Welty}, {Alsayyad}, {Kumar},
  {Lundgren}, {Shanidze}, {Smith}, {Vanlandingham}, {Baugher}, {Hall},
  {Jenkins}, {Menard}, {Rao}, {Tumlinson}, {Turnshek}, {Yip}, \&
  {Brinkmann}}]{ykv+06}
{York}, D.~G., {et~al.} 2006, \mnras, 367, 945

\bibitem[{{Zheng} {et~al.}(1997){Zheng}, {Kriss}, {Telfer}, {Grimes}, \&
  {Davidsen}}]{zheng1997}
{Zheng}, W., {Kriss}, G.~A., {Telfer}, R.~C., {Grimes}, J.~P., \& {Davidsen},
  A.~F. 1997, \apj, 475, 469

\end{thebibliography}

%\input{../Tables/observations_table.tex}
%\input {../Tables/hitable.tex} 
%\input{../Tables/dust_table.tex}

%%%%%%%%%%%%%%%%%%%%%%%%%%%%%%%%%%%%5

%%%%%%%%%%%%%%%%%%%%%%%%%%%%%%%%%%%%%%%%%%%%%%%%%%%%%%%%%%%%%%%%%%%%%%%%%%%%%%
%%%%%%%%%%%%%%%%%%%%%%%%%%%%%%%%%%%%%%%%%%%%%%%%%%%%%%%%%%%%%%%%%%%%%%%%%%%%%%
%Lya profile fits

\begin{center}
\includegraphics[width=6.8in]{f1a.eps}
\end{center}

\begin{center}
\includegraphics[width=6.8in]{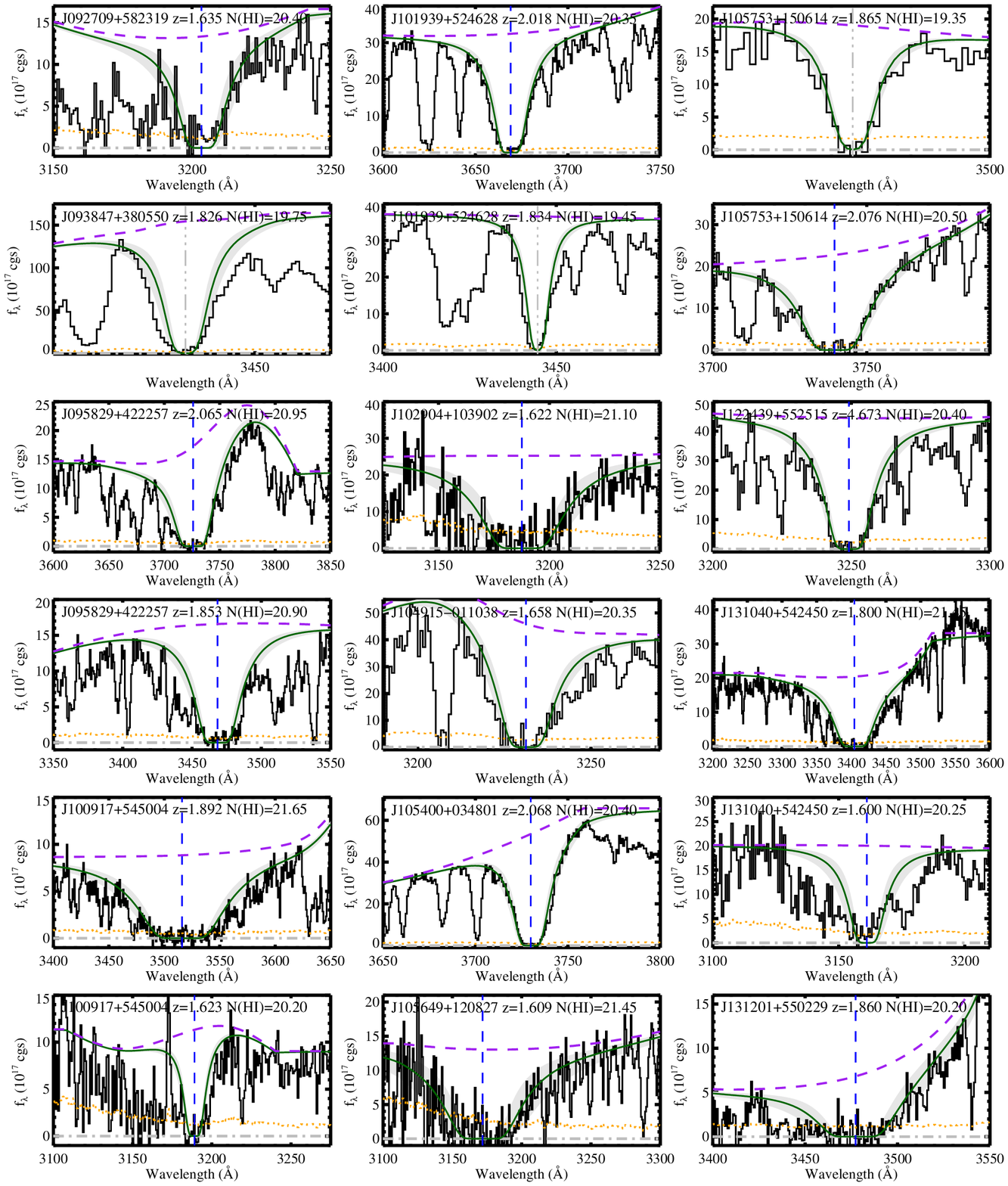}
\end{center}

\begin{center}
\includegraphics[width=6.8in]{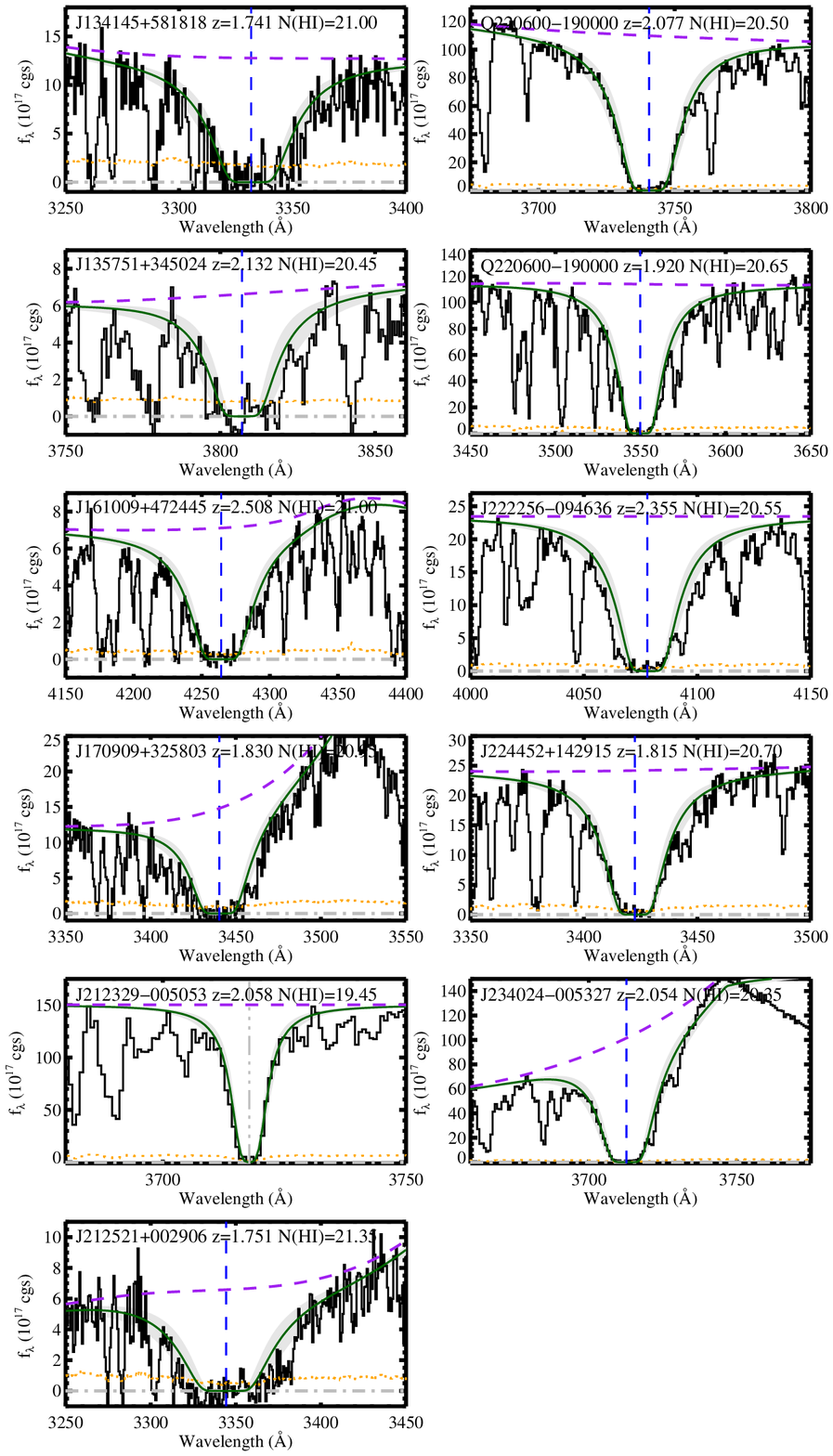}
\end{center}

\end{document}